\documentclass[aps,twocolumn,pra,superscriptaddress,floatfix,showpacs]{revtex4-1}
\usepackage{amsmath}
\usepackage{graphicx}
\usepackage{subfigure}
\usepackage[colorlinks=true,linktoc=page,linkcolor=blue,citecolor=blue,urlcolor=blue]{hyperref}
\usepackage{color}
\usepackage{epstopdf}
\usepackage{float}

\def\lsim{\mathrel{\mathpalette\gl@align<}}
\def\gsim{\mathrel{\mathpalette\gl@align>}}
\def\gl@align#1#2{\lower.6ex\vbox
{\baselineskip\z@skip\lineskip\z@
\ialign{$\m@th#1\hfil##\hfil$\crcr#2\crcr\sim\crcr}}}

\makeatother

\newcommand{\vect}[1]{\boldsymbol{\mathrm{#1}}}
\newcommand\ba{\begin{eqnarray}}
\newcommand\ea{\end{eqnarray}}
\newcommand\be{\begin{equation}}
\newcommand\ee{\end{equation}}

\usepackage[normalem]{ulem}
\usepackage{sidecap,tikz}
\DeclareRobustCommand{\orcidicon}{\hspace{-1.0mm}
	\begin{tikzpicture}
	\draw[lime, fill=lime] (0.0,0.0) 
	circle [radius=0.15] 
	node[white] {{\fontfamily{qag}\selectfont \tiny \,ID}};
	\draw[white, fill=white] (-0.0525,0.095) 
	circle [radius=0.007];
	\end{tikzpicture}
	\hspace{-3.0mm}
}
\foreach \x in {A, ..., Z}{\expandafter\xdef\csname orcid\x\endcsname{\noexpand\href{https://orcid.org/\csname orcidauthor\x\endcsname}
		{\noexpand\orcidicon}}
}

\begin{document}

\title{Anomaly in dynamical quantum phase transition in non-Hermitian system with extended gapless phases}

\author{Debashish Mondal}
\email{debashish.m@iopb.res.in}
\affiliation{Institute of Physics, Sachivalaya Marg, Bhubaneswar-751005, India}
\affiliation{Homi Bhabha National Institute, Training School Complex, Anushakti Nagar, Mumbai 400094, India}

\author{Tanay Nag\orcidB{}}
\email{tanay.nag@physics.uu.se}
\affiliation{Department of Physics and Astronomy, Uppsala University, Box 516, 75120 Uppsala, Sweden}


\begin{abstract}
The dynamical quantum phase transitions (DQPTs) and the associated winding numbers have been extensively studied in the context Hermitian system. We consider the non-Hermitian analogue of $p$-wave superconductor, supporting Hermitian gapless phase with complex hopping, in presence of on-site  or superconducting loss term. This
allows us to investigate the effect of non-Hermitian gapless phases on the DQPTs in addition to the Hermitian gapless phases. Our findings indicate that contour analysis of the underlying Hamiltonian, enclosing the origin and/or exceptional points, can predict the occurrences of DQPTs except the quench within the gapless phases. For the Hermitian case with initial and final Hamiltonians both being Hermitian,  we find  non-monotonic integer jump for the winding number as the  hallmark signature of the gapless phase there. For the hybrid case with initial and final Hamiltonians being Hermitian and non-Hermitian respectively, winding number exhibits integer spike in addition to the non-monotonic integer jumps. For the   non-Hermitian case  with initial and final Hamiltonians both being non-Hermitian, the winding number show half-integer jumps for lossy superconcuctivity that does not have any Hermitian analogue. On the other hand, the  integer jumps in winding number is observed for  lossy chemical potential. We understand  our findings by connecting them with the profile of Fisher zeros and number of exceptional points and/or origin.

\end{abstract}

\maketitle

\section{Introduction}

The thermodynamic limit is necessary to 
examine the equilibrium  phase transitions when the underlying system is well described by microscopic Hamiltonian without any singular interactions \cite{fisher1967theory}.  The  non-analyticities in the free-energy density, marked by the zeros of the partition function namely, Fisher zeros in complex temperature (magnetic field) plane,  lead to  temperature (magnetic field) driven liquid-gas (paramagnet-ferromagnet) transition  ~\cite{fisher1967theory,LYF1, LYF2}. The above idea is then generalized to dynamic quantum system where the non-analyticities in the dynamical free-energy density  turns up to be instrumental in predicting the dynamical quantum phase transitions (DQPTs) in complex time plane \cite{heyl13,PhysRevB.87.195104,PhysRevB.90.125106,PhysRevLett.113.265702,PhysRevLett.115.140602,Heyl_2018,Bhattacharya17,Jafari19a,Uhrich20}. The DQPTs are considered to be the dynamical analogs to equilibrium quantum
phase transitions when the initial state is orthogonal to the time evolved state under a sudden quench.
Interestingly, the DQPTs are not always intimately connected to the 
sudden quench across the quantum critical point (QCP)  \cite{Vajna14,Schmitt15,Halimeh17,Silva18,Halimeh20c,Hashizume22,Lang18,Homrighausen17,rossi2022non,mishra2020disordered}. The slow quench is also found to exhibit DQPTs \cite{SS,PhysRevB.92.104306,Divakaran16,Dutta17}. Importantly, analogous to order
parameters for the conventional phase transitions, the dynamical topological order parameter
is introduced to characterize the topological properties of the
real-time dynamics through winding number \cite{PhysRevB.91.155127,Budich1}. The  realm of DQPTs is extended from free fermion models \cite{PhysRevB.89.161105} to interacting \cite{PhysRevB.92.235433,PhysRevB.89.125120,Halimeh17,Modak21} as well as bosonic systems \cite{Abdi19,PhysRevB.103.064306,PhysRevResearch.4.013002}. In the context of Floquet driving, Fisher zeros exhibit intriguing profile \cite{Zamani20,Jafari22,Jafari21a}.
The DQPTs are experimentally observed in trapped-ion  \cite{PhysRevLett.119.080501}, nuclear magnetic resonance \cite{Nie20}, optical lattice \cite{flaschner2018observation} systems. Notice that DQPT is essentially related to Loschmidt amplitude (LA) which is extensively studied in the context of quantum information theoretic measures namely, decoherence \cite{Nag12,Sachdeva14,Nag16,Suzuki16,PhysRevLett.96.140604,Cucchietti03,Jafari17b}. \textcolor{black}{Notice that DQPTs are investigated in the context of time crystals \cite{Kosior18,Kosior18b}, and  Floquet driving \cite{Zamani20,Jafari22,Jafari21,Jafari17}. Importantly, non-decaying nature of LA in the Floquet DQPT enables a better handle to probe the phenomena experimentally \cite{zhou2021floquet,Yang19}.
}


Very recently, non-Hermitian analogue of the underlying Hermitian quantum systems have received enormous attention due to their vast applicability in open quantum system \cite{Bergholtz19,Yang21}, quasiparticles system with finite lifetime \cite{kozii2017non,Yoshida18,Shen18} as well as their
practical realizations in meta-materials such as  cold atom \cite{Gou20,li2019observation}, photonic \cite{Zeuner15,weimann2017topologically} and acoustic \cite{Weiwei18,Gao20} systems. The non-Hermitian system is found to host exceptional points (EPs)
where eigenstates, corresponding to degenerate bands,
coalesce \cite{Bergholtz21,ghatak2019new,ashida2020non,Kawabata19,Shen18}. The time evolution becomes non-unitary  for the non-Hermitian case that can non-trivially modify the emergence of DQPT ~\cite{Zhou1,Zhou_2021,PhysRevA.105.022220,Hamazaki2021}. As a result, the time dependent winding number is expected to show intriguing jump profile for quenching across the EPs. Given the fact that the non-Hermitian system is far more less explored in the context of DQPT for gapless systems \cite{PhysRevB.92.075114,Jafari2019}, we seek answers for the following questions: what are the roles of gapless regions in Hermitian and non-Hermitian systems, bounded by Hermitian  critical lines and EPs respectively, for DQPT and correspondingly in the subsequent evolution for the  winding number?
How does the profile of Fisher zeros change with non-Hermiticities?


Considering a variant of one-dimensional (1D) $p$-wave superconductor Kitaev chain with 
complex hopping and non-Hermitian terms, we examine the emergence of DQPT through various paths for sudden quenching from gap to gap and gapless as well as gapless to gapless regions (see Fig.~\ref{fig:1}). We find that contour analysis, based on the inclusion of origin (EPs) for underlying Hermitian (non-Hermitian) Hamiltonians and their chiralities, can correctly indicate the   non-analyticities in the rate function in all the above quench protocol except the quenching within the gapless phase [see Figs.~\ref{fig:2} \ref{fig:5}, \ref{fig:8}, and \ref{fig:11}]. In the Hermitian case, the non-monotonic behavior of winding number, associated with the DQPTs, is a marked signature for this gapless phase (see Fig.~\ref{fig:6}). For hybrid case with initial Hermitian and final non-Hermitian Hamiltonians, the contour analysis can predict DQPTs expect  when the final Hamiltonian resides in the gapless non-Hermitian phase. The winding number demonstrates an integer spike  in addition to the non-monotonic jumps for the quench discussed above (see Fig.~\ref{fig:10_2}). For the non-Hermitian case, we remarkably find DQPTs, associated with half-integer  jumps besides the non-monotonic integer jumps in the winding number, exist only for quenching within the gapless phases in presence of lossy superconductivity (see Fig.~\ref{fig:14}). Under lossy chemical potential, the DQPTs are observed as well. We understand the behavior of winding number by analyzing the number of enclosed EPs and origin within the appropriate contours as well as 
the structure of Fisher zeros. The critical momentum and time, causing the DQPTs, are estimated from the closed form expressions for both the Hermitian and non-Hermitian Hamiltonians.

The paper is organized as follows. The Sec.~\ref{model} describes the model and the framework of DQPTs. We analyse the phase diagrams under non-Hermiticities. We examine the results in Sec.~\ref{results} where we explore the contour profiles, Fisher zeros, DQPTs, geometric phases and winding numbers.  The Hermitian (Sec.~\ref{results1}), hybrid (Sec.~\ref{results2}) and non-Hermitian (Sec.~\ref{results3}) cases are 
demonstrated when the underlying Hamiltonians are both Hermitian, Hermitian as well as non-Hermitian and both non-Hermitian, respectively. We conclude in Sec.~\ref{conclusion}.        

\section{Model and Dynamical Quantum Phase Transitions}
\label{model}

We consider 1D $p$-wave superconductor with complex hopping as follows $H(\gamma_1,\gamma_2,\phi)= \sum_{k} \psi_k {\cal H}_{k}(\gamma_1,\gamma_2,\phi) \psi^{\dagger}_k  $
\cite{DeGo1,Manisha1,Rajak1} 
\begin{eqnarray}
&&{\cal H}_{k}(\gamma_1,\gamma_2,\phi) = 2w_0 \sin\phi \sin k~I 
+\Big(2 \Delta \sin k +\frac{i\gamma_2}{2}\Big)\sigma_{y} \nonumber \\
&&- \Big(2 w_0 \cos\phi \cos k +\mu  
+  \frac{i\gamma_1}{2}\Big)\sigma_{z}  =\vect{h}_k \cdot \vect{\sigma}
\label{eq:Momentum_Ham}.
\end{eqnarray}
where $w_0$, $\phi \in [0,\pi/2]$, $\mu$, and  $\Delta$ are the nearest neighbour hopping amplitude, phase of the hopping amplitude, chemical potential and superconducting gap respectively, with $\vect{h}_k=\{h^0_k,h^y_k,h^z_k\}$ and $\vect{\sigma}=\{\sigma_0,\sigma_y,\sigma_z\}$. The Hamiltonian is written in the basis of $\psi_k= (c_k, c^{\dagger}_{-k})$ using pseudo-spin degrees
of freedom formed by the fermion particle-hole subspace. 
The energy of the Hamiltonian ${\cal H}_{k}$ is found to be $h^{\pm}_k=h^0_k \pm \sqrt{(h^y_k)^2 + (h^z_k)^2} $.
Here, $\gamma_{1,2}$ represent the non-Hermitian factors. $\gamma_1$ is associated with the on-site gain and loss terms that can be caused by the imaginary part of the self energy for open  and/or interacting quantum systems. We refer to the above instance as the lossy chemical potential. On the other hand,
$\gamma_2$ is modeled like a non-reciprocal effects in the gap function which does not have any practical analogue so far to the best of our knowledge.  We indicate this situation as the lossy superconductivity.

The Hermitian analogue of the model preserves particle-hole symmetry (PHS) ${\mathcal P} {\cal H}_{k}(0,0,\phi){\mathcal P}^{-1}=-{\cal H}_{-k}(0,0,\phi)$ with ${\mathcal P}=\sigma_x {\mathcal K}$ (${\mathcal K}$ denotes the complex conjugation) while time reversal symmetry (TRS) ${\mathcal T}={\mathcal K}$ and chiral symmetry (CS) ${\mathcal C}={\mathcal T}{\mathcal P}$ are preserved only for real hopping: ${\mathcal T} {\cal H}_{k}(0,0,0){\mathcal T}^{-1}={\cal H}_{-k}(0,0,0)$ and ${\mathcal C} {\cal H}_{k}(0,0,0){\mathcal C}^{-1}=-{\cal H}_{k}(0,0,0)$. On the other hand, for non-Hermitian case $\gamma_1\ne 0,\gamma_2=0$, only PHS continues to be preserved  ${\mathcal P}_{-} {\cal H}^T_{k}(\gamma_1,0,\phi){\mathcal P}^{-1}_{-}=-{\cal H}_{-k}(0,0,\phi)$ with ${\mathcal P}_{-}=\sigma_x$ while ${\cal H}_{k}(\gamma_1,0,0)$ respects  TRS$^{\dagger}$  ${\mathcal T}_{+}=\sigma_0$ and CS ${\mathcal C}_{-}=\sigma_x$: ${\mathcal T}_{+} {\cal H}^T_{k}(\gamma_1,0,0){\mathcal T}^{-1}_{+} = {\cal H}_{-k}(\gamma_1,0,0)$ and ${\mathcal C}_{-} {\cal H}_{k}(\gamma_1,0,0){\mathcal C}^{-1}_{-}= -{\cal H}_{k}(\gamma_1,0,0)$. Interestingly, we find $\Gamma {\cal H}_{k}(\gamma_1,0,0)\Gamma^{-1}= {\cal H}_{-k}(\gamma_1,0,0)$ where the inversion symmetry is denoted by $\Gamma=\sigma_z$. In the other non-Hermitian case $\gamma_1= 0,\gamma_2 \ne 0$, the CS is only respected for ${\cal H}_{k}(0,\gamma_2,0)$ while TRS, PHS and IS are broken for ${\cal H}_{k}(0,\gamma_2,0)$ and ${\cal H}_{k}(0,\gamma_2,\phi)$. Therefore, ${\cal H}_{k}(0,\gamma_2,\phi)$  is maximally symmetry broken where PHS is only preserved  for ${\cal H}_{k}(\gamma_1,0,\phi)$.

\begin{figure}[H] 
\includegraphics[trim=0.0cm 0.75cm 0.2cm 0.2cm, clip=true, height=!,width=1.0\columnwidth]{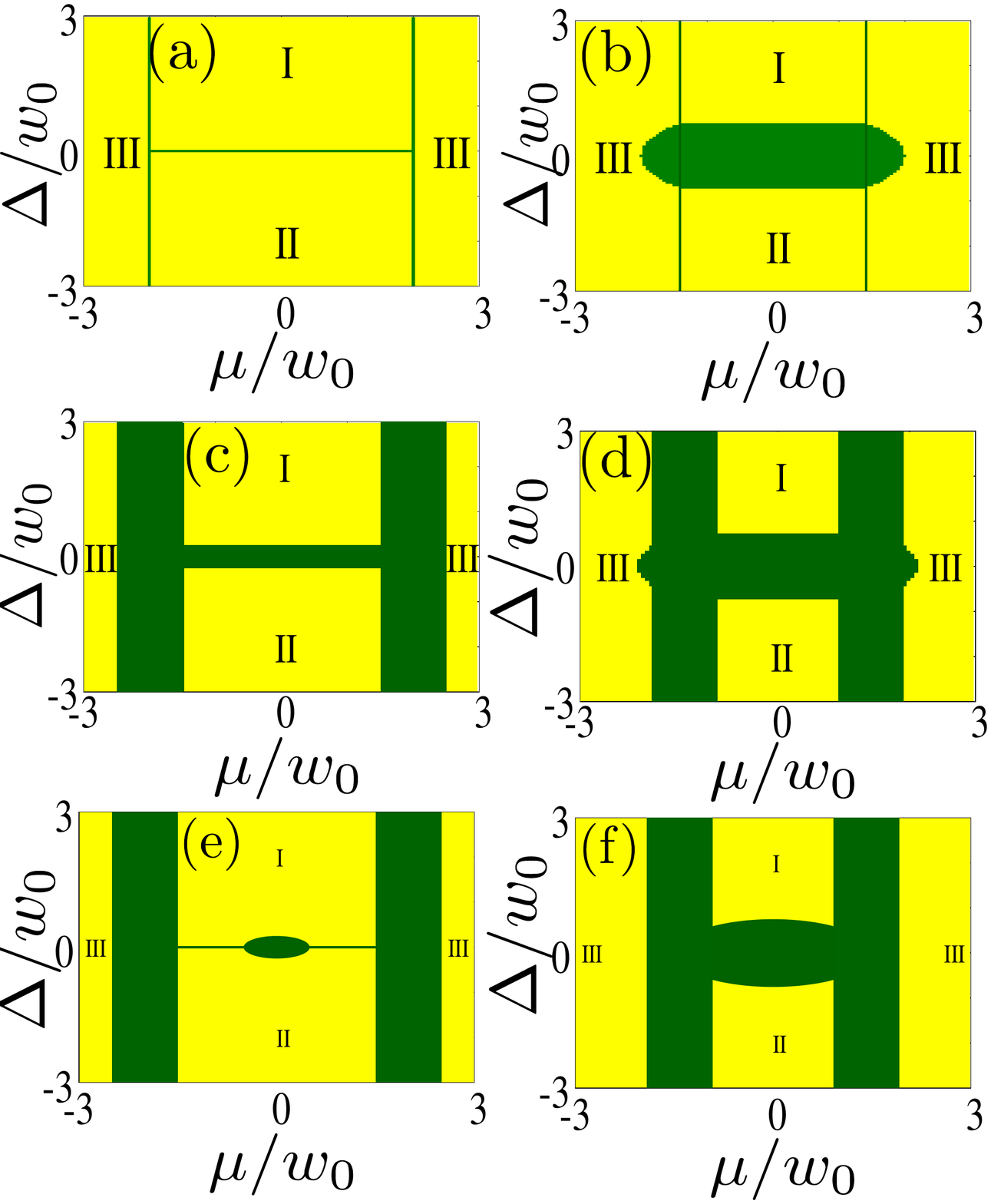}
\caption{The phase diagram of the model Hamiltonian Eq.~(\ref{eq:Momentum_Ham}) for $H(0,0,\phi)$ in (a), (b), $H(0,\gamma_2,\phi)$ in  (c), (d) and $H(\gamma_1,0,\phi)$ in  
(e) and (f). We consider $\phi=0$ $[\pi/4]$ in (a), (c), and (e) [(b), (d), and (f)]. The  yellow region represents  the gapped  phase I, II and III  while green regions are gapless. We consider $\gamma_1=\gamma_2=1$. } \label{fig:1}
\end{figure}

Notice that the Hamiltonian (\ref{eq:Momentum_Ham}) reduces to Kitaev model for  1D $p$-wave superconductor, supporting two topological phases for $|\mu|< 2 w_0$ 
trivial phases for $|\mu|>2w_0$,  when  $\phi=\gamma_{1,2}=0$ \cite{kitaev2001unpaired}. The Hamiltonian Eq.~(\ref{eq:Momentum_Ham})  becomes gapless for  critical momentum $k_*$ when the real part satisfies the following  condition  
\begin{eqnarray}
&&\Big(2w_0 \cos {\phi} \cos k_{*} +\mu \Big)^{2} -\frac{\gamma^2_1}{4} +4 \Delta^2 \sin^2 k_{*} -\frac{\gamma^2_2}{4} \nonumber \\
&&=4 w^2_0 \sin^2 {\phi} \sin^2 k_{*}
\label{eq:non_hermi_kc}
\end{eqnarray} 
Starting with the simple Hermitian case $H(0,0,0<\phi<\pi/2)$,  one can find  a rectangular gapless region, bounded vertically (horizontally)  by $\Delta =\pm w_{0} \sin \phi$ ($\mu=\pm 2 w_{0} \cos\phi$), that separates topological I phase from topological II phase (topological I and II phases from non-topological III phase). 
This is shown in Fig. \ref{fig:1} (a) and (b) for $H(0,0,\phi=0)$ and $H(0,0,\phi=\pi/4)$, respectively.
Two elliptical arcs both in left and right sides of the rectangle region appear for infinitesimal value of $\phi$. Importantly, within the whole gapless green region, there exist a 
set of values  for  $k_c$ satisfying the above gapless condition
in Eq.~(\ref{eq:non_hermi_kc}) with $\gamma_1=\gamma_2=0$.
For the non-Hermitian case,  we refer to the yellow (green) region as the gapped I, II and III (gapless) instead of topological and non-topological phases without loss of generality.   
We consider two distinct situations $\gamma_1\ne 0,\gamma_2=0$
and $\gamma_1=0, \gamma_2\ne 0$ for the non-Hermitian case and 
$\gamma_1=\gamma_2=0$ for the Hermitian case. 

The EPs arising in the $h_y$-$h_z$ plane for the non-Hermitian case are given by ${\rm Im}[h^y_k]=\pm{\rm Re}[h^z_k]$ and ${\rm Im}[h^z_k]=\mp{\rm Re}[h^y_k]$. 
The  rectangle-like region  is  bounded by horizontal lines
$\Delta =\pm \sqrt{w^2_{0} \sin^2 \phi + \gamma_2^2/16}$ [convex lines $\Delta =\pm \sqrt{w^2_{0} \sin^2 \phi + \gamma_1^2/16-\mu^2/4}$] for $\gamma_1=0,\gamma_2\ne 0$ [$\gamma_1 \ne 0,\gamma_2= 0$] with $k_*=\pm \pi/2$. On the other hand, the vertical gapless region, bounded by $\mu=\pm 2 w_{0} \cos\phi \pm {\gamma_2}/{2}$ [$\mu=\pm 2 w_{0} \cos\phi \pm {\gamma_1}/{2}$] with $k_*=0,\pi$  for $\gamma_1=0,\gamma_2\ne 0$ [$\gamma_1 \ne 0,\gamma_2= 0$].  Therefore, the vertical gapless region of width $\gamma$, enclosed between $ 2 w_{0} \cos\phi - {\gamma}/{2} < \mu <2 w_{0} \cos\phi + {\gamma}/{2}$ [$ -2 w_{0} \cos\phi - {\gamma}/{2} < \mu <-2 w_{0} \cos\phi + {\gamma}/{2}$]
for $\mu>0$ [$\mu<0$] is an intriguing outcome of non-Hermiticity.
The above features are demonstrated in Fig. \ref{fig:1} (c), (d) and (e), (f) for $H(0,\gamma,\phi=0,\pi/4)$, and $H(\gamma,0,\phi=0,\pi/4)$, respectively.
This is contrast to  the Hermitian case where the rectangular region gets modified due to non-Hermiticity irrespective of the any particular choice discussed in the present case. Interestingly, for $\phi=\pi/2$,
the topological phases disappear completely and the gapless region is bounded by an ellipse $4 \Delta^2 + \mu^2 = 4w^2_0 +\gamma^2/4$ with $\gamma_1=0,\gamma_2=\gamma\ne 0$ or $\gamma_2=0,\gamma_1=\gamma\ne 0$.


The LA accounts for the overlap between the initial state $|\Psi_i\rangle$ and the time-evolved state 
$|\Psi(t)\rangle=e^{-i H_f t}|\Psi_i\rangle$ is 
found to be $G(t)=\langle \Psi_i|e^{-i H_f t}|\Psi_i\rangle$ 
when the system undergoes a sudden quench from initial  Hamiltonian $H_i$ to final $H_f$ \cite{heyl13,Budich1}. The LA can be expressed as $G(t)=\Pi_k g_k(t)$ ~\cite{Zhou1}
\begin{equation}
 g_k(t)=\cos(h_{k,f}t)-i\sin(h_{k,f}t)\langle\psi_{k,i}|\frac{{\cal H}_{k,f}}{h_{k,f}}|\psi_{k,i}\rangle \label{eq:g_k},
\end{equation}
where initial and final Hamiltonian are expressed in terms of individual momentum modes: ${\cal H}_{k,i(f)} |\psi_{k,i(f)}\rangle=h_{k,i(f)} |\psi_{k,i(f)}\rangle $. 
In the thermodynamic limit the rate function, associated with LA, is given by
\begin{equation} 
I(t)=-\frac{1}{2\pi}\int_{0}^{2\pi}dk\ln|{g}_{k}(t)|^{2}.
\label{eq:rt}
\end{equation} 
Interestingly, when  the argument of ``ln"  in Eq.~(\ref{eq:rt})  becomes zero,
the DQPTs occur. The above criterion leads to  the Fisher zeros~\cite{fisher1967theory,LYF1,LYF2}. This happens at a few imaginary times $z=it$ as follows
\begin{equation}
z_{n,k}=i\frac{\pi}{h_{k,f}}\left(n+\frac{1}{2}\right)+\frac{1}{h_{k,f}}{\rm arctanh}\langle\psi_{k,i}|\frac{{\cal H}_{k,f}}{h_{k,f}}|\psi_{k,i}\rangle,\label{eq:LYF}
\end{equation}
with certain momenta $k$ referred to as the critical momenta $k_c$. 
Here $n$ denotes the integer  numbers. Note that the identity term in Hamiltonian Eq.~(\ref{eq:Momentum_Ham}) do not change the LA qualitatively under non-equilibrium dynamics and hence we consider ${\cal H}_{k,i(f)}=h^y_{k,i(f)} \sigma_y + h^z_{k,i(f)} \sigma_z$
and $h^{\pm}_{k,i(f)}=\pm \sqrt{\big(h^y_{k,i(f)}\big)^2 + \big(h^z_{k,i(f)}\big)^2}$ for further calculations .

For general non-Hermitian case with $\gamma_{1,2}\ne 0$,  $h_{k,f}$ and
${\rm arctanh}\langle\psi_{k,i}|\frac{{\cal H}_{k,f}}{h_{k,f}}|\psi_{k,i}\rangle$ both can in general be complex.
Notice that $|\psi_{k,i(f)}\rangle $ and $\langle\psi_{k,i(f)}|$ represent the right and left eigenvectors of ${\cal H}_{k,i(f)}$
with $\sum_n|\psi^n_{k,i(f)}\rangle \langle\psi^n_{k,i(f)}|=I$ and 
$\langle\psi^n_{k,i(f)}| \psi^m_{k,i(f)}\rangle=\delta_{mn}$ owing to bi-orthogonalization.
As a result, $z_{n,k}$ receives finite real contribution from both the above terms. A careful analysis suggests that, provided ${\rm Re}[z_{n,k_c}]=0$, the critical momenta $k_{c}$ satisfies the following condition  
\begin{equation}
\pi (n + \frac{1}{2}){\rm Im}[h_{k_c,f}]+ {\rm Re}[h_{k_c,f}]{\rm Re}[A_{k_c}] + {\rm Im}[h_{k_c,f}]{\rm Im}[A_{k_c}]=0 
\label{nh_kc}
\end{equation}
with $A_{k_c}={\rm arctanh}\langle\psi_{k_c,i}|\frac{{\cal H}_{k_c,f}}{h_{k_c,f}}|\psi_{k_c,i}\rangle$. Interestingly, one can find multiple critical momenta in the non-Hermitian case in contrast to the Hermitian case as discussed below. 
The critical time $t_c$ corresponding to the above $k_c$ is given by the imaginary part of $z_{n,k}$ as follows
\begin{eqnarray}
t_c&=& \pi (n + \frac{1}{2})\frac{{\rm Re}[h_{k_c,f}]}{|h_{k_c,f}|^2} \nonumber \\
&+& \frac{{\rm Re}[h_{k_c,f}]{\rm Im}[A_{k_c}] - {\rm Im}[h_{k_c,f}]{\rm Re}[A_{k_c}]}{|h_{k_c,f}|^2} 
\label{nh_tc}
\end{eqnarray}

The time-dependent winding number further characterizes the non-analytic behavior in the rate-function Eq.~(\ref{eq:rt}) at real critical time $t_c$   
\begin{equation}
\nu(t)=\frac{1}{2\pi}\int_{0}^{2\pi}dk\left[\partial_{k}\phi_{k}^{{\rm G}}(t)\right].\label{eq:WN}
\end{equation}
Here the geometric phase  of the return amplitude is given by $\phi_k^G(t)=\tilde {\phi}_k(t)-\phi_k^{\rm dyn}(t)$, with total phase  $\tilde{\phi}_k(t)= - i \ln \left[\frac{g_k(t)}{|g_k(t)|}\right]$ and the dynamical phase is found to be ~\cite{Zhou1,Gong1}
\begin{eqnarray}
\phi_{k}^{{\rm dyn}}(t)=  -\int_{0}^{t}ds\frac{\langle\psi_{k,i}(s)|{\cal H}_{k,f}|\psi_{k,i}(s)\rangle}{\langle\psi_{k,i}(s)|\psi_{k,i}(s)\rangle} \nonumber \\
+\frac{i}{2}\ln\left[\frac{\langle\psi_{k,i}(t)|\psi_{k,i}(t)\rangle}{\langle\psi_{k,i}(0)|\psi_{k,i}(0)\rangle}\right],\label{eq:dyn} 
\end{eqnarray}
where $|\psi_{k,i}(t)\rangle=e^{-i{\cal H}_{k,f} t }|\psi_{k,i}\rangle$ and $\langle\psi_{k,i}(t)|=\langle\psi_{k,i}| e^{i {\cal  H^\dagger}_{k,f} t}$ are the time-evolved  right and left eigenvectors of ${\cal H}_{k,i}$ respectively. Note that the
upper limits of the integration in Eq.~(\ref{eq:rt}) and Eq.~(\ref{eq:WN}) can be $\pi$ instead of $2\pi$, once the eigenvalue spectrum of  ${\cal H}_{k,i(f)}$
becomes $\pi$-periodic with respect to $k$.

Having discussed the DQPT in the context of non-Hermitian Hamiltonian, we now briefly mention the Hermitian analogue i.e., $\gamma_{1,2}=0$. To this end, one can consider
$\langle\psi_{k,i(f)}|=(\cos (\theta_{k,i(f)}/2), \sin (\theta_{k,i(f)}/2))$ and  $|\psi_{k,i(f)}\rangle=\langle\psi_{k,i(f)}|^{T}$ without loss of generality as the ground state for $2$-level system ${\cal H}_{k,i(f)}$, corresponding to energy $h_{k,i(f)}=-\sqrt{[h^y_{k,i(f)}]^{2} + 
[h^{z}_{k,i(f)}]^2}$  where $\theta_{k,i(f)}= {\rm arctan}[h^y_{k,i(f)}/h^{z}_{k,i(f)}]$. 
The LA, derived from Eq.~(\ref{eq:g_k}), is thus found to be  $
g_{k}(t)=\cos^{2}\phi_{k} + \sin^{2} \phi_{k}~ \exp[- 2i h_{k,f} t]$ where  $\phi_{k}=(\theta_{k,i} -\theta_{k,f})/2$.
Following the same line of argument as presented in Eq.~(\ref{eq:LYF}),
the Fisher zeros acquires the form $z_{n,k} =i\pi(n+1/2)/h_{k,f}+ \ln(\tan^{2}\phi_{k})/2h_{k,f}$.
The DQPTs occurs at $t_c=\pi(n+1/2)/h_{k_c,f}$ corresponding to critical momentum $k_c$ for which $\tan^{2}\phi_{k_c}=1$. Unlike the non-Hermitian case, the $k_c$ is uniquely determined by $|\cos \phi_{k_c}|=|\sin\phi_{k_c}|=1/\sqrt{2}$ such that
$\tan \theta_{k_c,i}=-\cot\theta_{k_c,f}$ i.e., 
\begin{equation}
\frac{h_{k_c,i}^{y} \times h_{k_c,f}^{y}}{h_{k_c,i}^{z} \times h_{k_c,f}^{z}}=-1\label{eq:k_c}.
\end{equation}
The winding number for Hermitian case, as demonstrated in Eq.~(\ref{eq:WN}), is determined by the geometric phase $\phi_k^G(t)=\tilde {\phi}_k(t)-\phi_k^{\rm dyn}(t)$
with the dynamical phase $\phi_k^{\rm dyn}(t)=-2h_{k,f} t \sin^2\phi_{k}$ and total phase \cite{Budich1,SS}
\begin{equation}
\tilde {\phi}_k(t) = \arctan \left[\frac{- \sin^2\phi_k \sin(2 h_{k,f} t)}{\cos^2\phi_{k} +\sin^2\phi_{k} \cos(2 h_{k,f} t)}\right] \label{eq:h2_phase}.
\end{equation}

Note that for quench involving non-Hermitian Hamiltonian,  we computed a renormalized rate function $I(t)\equiv-\frac{1}{2\pi}\int^{2\pi}_0 dk \ln[|{g}_{k}(t)|^2/|e^{-i{\cal H}^{f}_{k} t}|\psi^{i-}_{k}\rangle|^2]$ that allows us  to show the non-analytic behavior of $I(t)$  more clearly. This renormalized rate function does not affect the behavior of DQPTs  qualitatively  \cite{Zhou1}. \textcolor{black}{Here, $|\psi^{i-}_{k}\rangle$ is the ground state of the initial Hamiltonian.}


\section{Results}
\label{results}

We here focus on the emergence of  DQPTs for sudden quench involving gapless phase in Hermitian, and non-Hermitian cases. 
We consider $\phi=\pi/4$ and $w_0=1$ unless it is specified otherwise.

\subsection{Hermitian case:}
\label{results1}
\begin{figure}[H]\includegraphics[trim=1cm 5.9cm 0.8cm 4.5cm, clip=true, height=!,width=1\columnwidth]{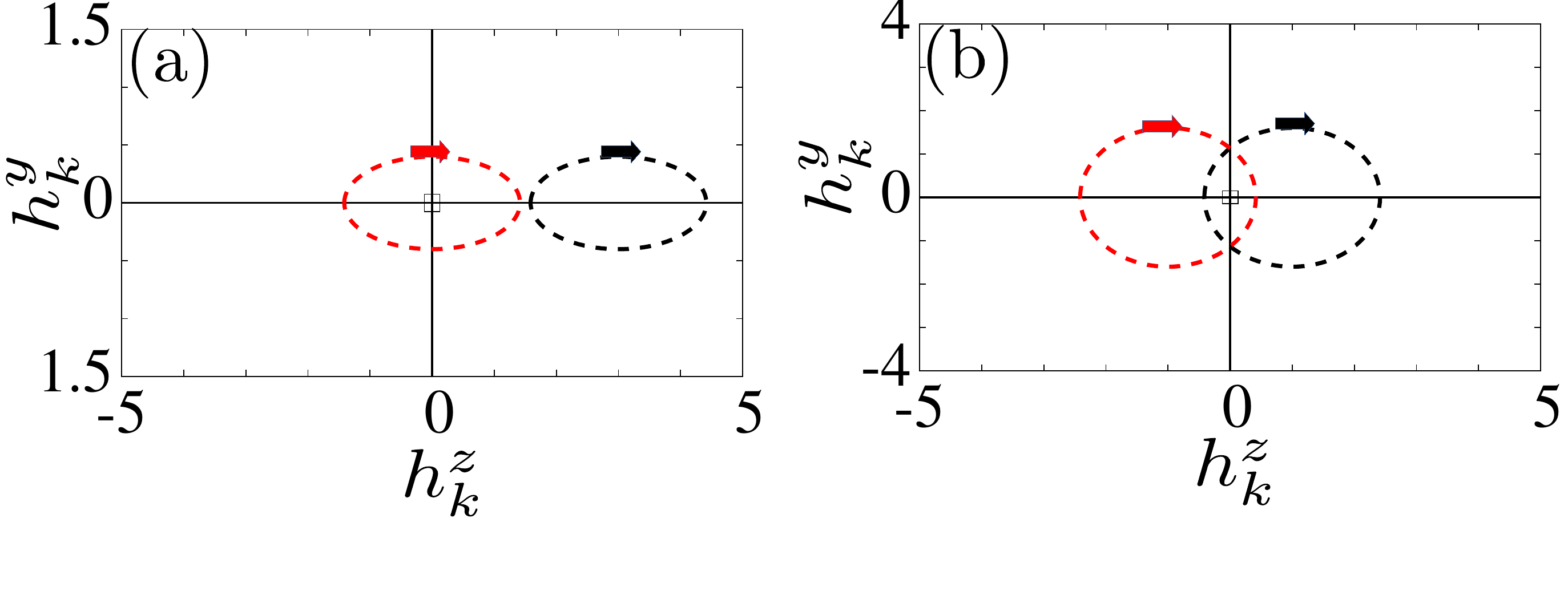}
	\caption{The parametric profiles of $h^y_k$ and $h^z_k$ for Hamiltonian Eq.~(\ref{eq:Momentum_Ham}) ${\cal H}_{k}(0,0,\pi/4)$ following the sudden quench across 
	the QCP from initial (marked by black dashed line) gapped phase to final (marked by red dashed line) gapless (a) and gapped (b) phases.  The arrows indicate the chirality of the contours as $k$ increases from $0$ to $2\pi$. The hollow square at the origin, designating QCP with $h^y_k=h^z_k=0$ such that $\Delta {h}_k= {h}^+_k- {h}^-_k=0$, is only enclosed when the Hamiltonian supports topological phase. The set of parameters considered for (a)  [(b)] is $(\Delta_i,\Delta_f,\mu_i,\mu_f)=(0.2,0.2,-3,0)$ [$(0.8,0.8,-1,1)$], suggesting the fact that only final [both initial and final] phase [phases] is [are] topological. 
	}\label{fig:2}
\end{figure}

The topological (non-topological) phase for $|\mu|<2w_0\cos \phi$ ($|\mu|>2w_0\cos \phi$) corresponds to a situation   
when the contours, defined by the path going around the BZ, of the gap terms $h^{y,z}_{k}$ do (do not) enclose the origin $ {h}_k=\{h^y_k,h^z_k\}=0$ point. The interesting point to note is that the contour of $h_k$ continues enclosing the origin irrespective of the gapless nature of the phase as long as $|\mu|<2w_0\cos \phi$. Therefore, the contours of $h_{k}$ is not sensitive to the gapless phase boundaries $\Delta =\pm w_0\sin \phi$.  
The chirality of the contours is determined by the flow of the momentum modes from $0$ to $2\pi$ in the $h^{y}_{k}$-$h^{z}_{k}$ parameter space. The chirality changes sign when $\Delta$ crosses zero from positive (negative) to negative (positive) values.
The contours for $ {h}_{k,i}$
and ${h}_{k,f}$ are demonstrated in Fig.~\ref{fig:2}  with black and red dashed lines, respectively.

\begin{figure}[H]\includegraphics[trim=0.1cm 0.9cm 0.4cm 0.4cm, clip=true, height=!,width=1\columnwidth]{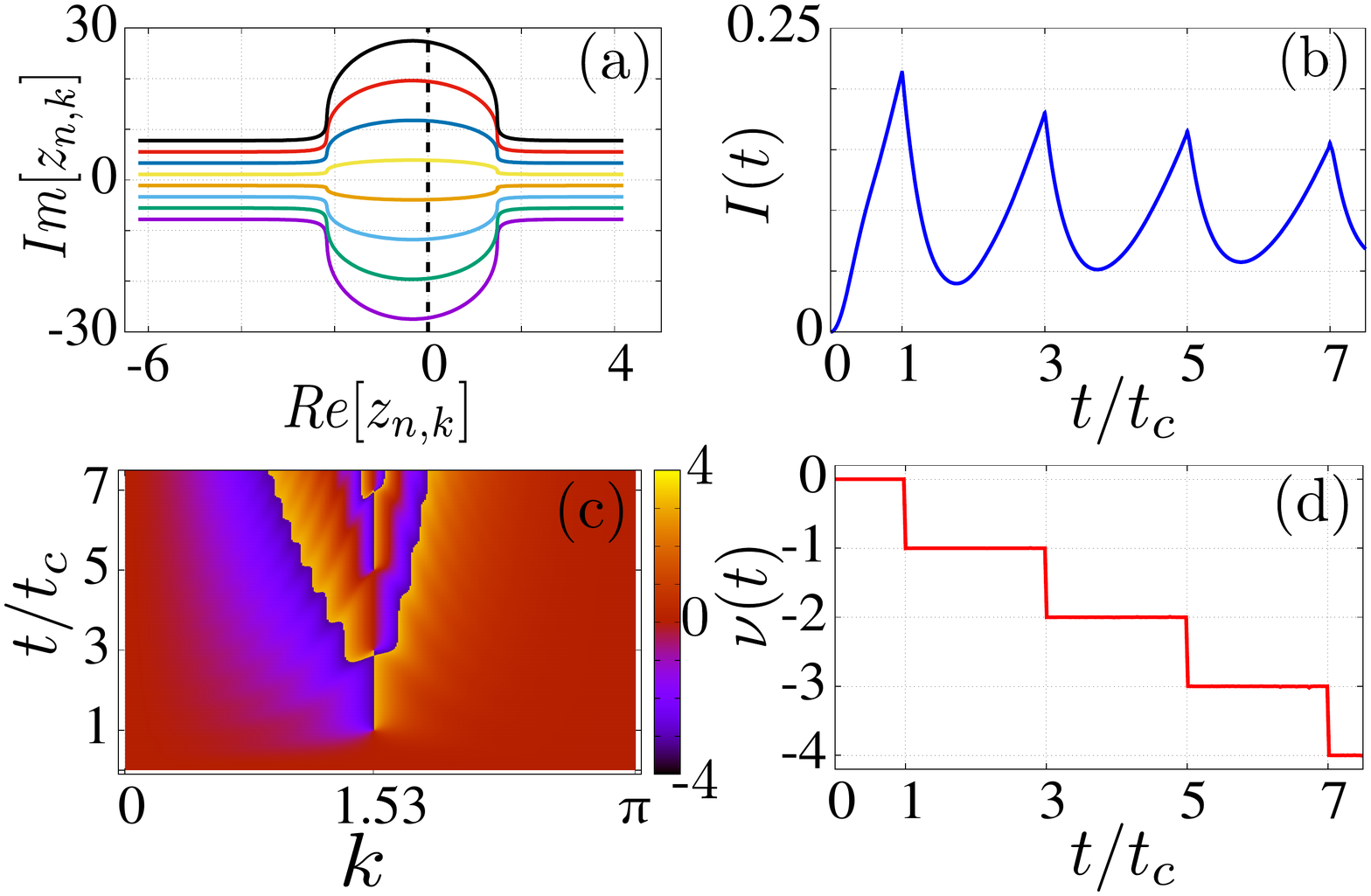}
	\caption{(a) The lines of Fisher zeroes $z_{n,k}$ with $n=-4$ (violet), $\cdots, 3$ (black), computed from Eq.~(\ref{eq:LYF}) for the case discussed in Fig.~\ref{fig:2} (a),  cross the imaginary axis suggesting the occurrence of DQPT. (b) Non-analytic behavior  in the rate function $I(t)$, obtained from Eq.~(\ref{eq:rt}), is clearly visible in time  at $t/t_{c}=1,3,5,7, \cdots$ where $t_c \approx 3.89$. (c)  The geometric phase $\phi_{k}^{G}(t)$, following Eq.~(\ref{eq:h2_phase}) and $\phi_k^{\rm dyn}(t)$,  changes  abruptly  around critical momentum $k_{c}\approx 1.53$ for $t/t_{c}=1,3,5,7, \cdots$ as
	shown in the $k$-$t$ plane. (d) The monotonic evolution of topological winding number $\nu(t)$, estimated from Eq.~(\ref{eq:WN}), exhibit unit jump   at $t/t_{c}=1,3,5,7, \cdots$ consistent with the evolution of $I(t)$.  }\label{fig:4}
\end{figure}

\begin{figure}[H]\includegraphics[trim=0.1cm 4.9cm 0.4cm 4.4cm, clip=true, height=!,width=1\columnwidth]{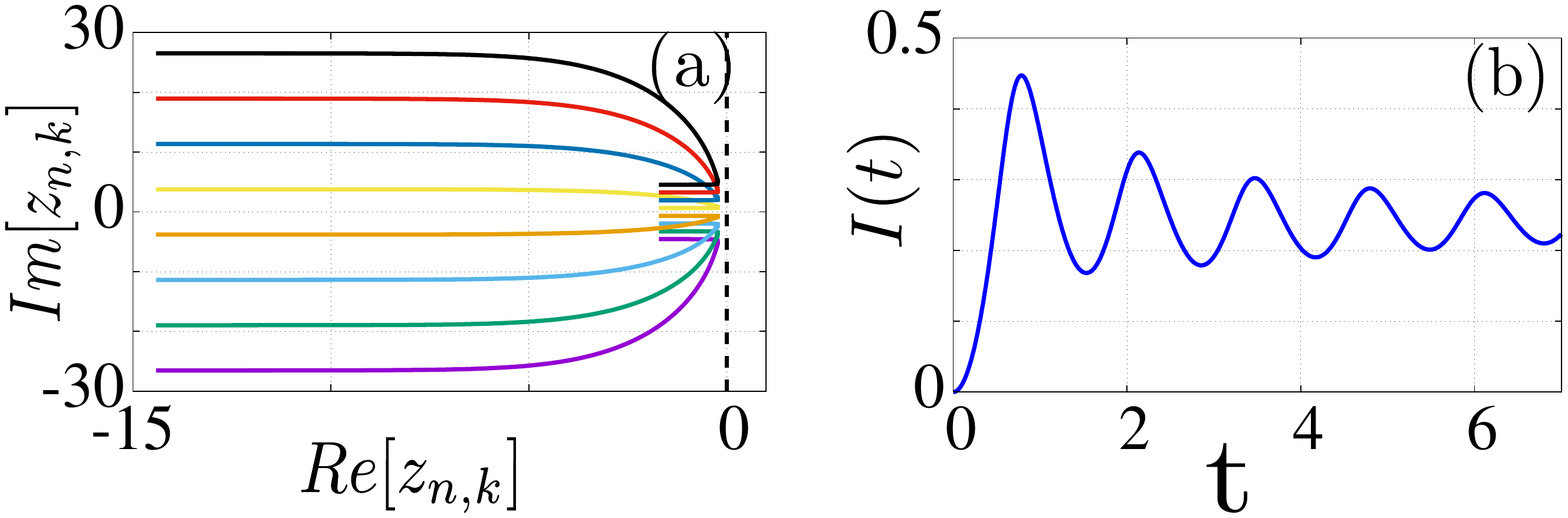}
	\caption{We repeat Fig.~\ref{fig:4} (a) and (b) for the case demonstrated in Fig.~\ref{fig:2} (b). (a) The Fisher zeros do not cross the imaginary axis. (b) The rate function does not encounter any singularities. The DQPTs are no longer observed even though the both initial and final Hamiltonians  both enclose the origin. }\label{fig:4_2}
\end{figure}

First, we consider the quench $\mu_i \to \mu_f$ keeping  fixed $|\Delta|<w_0\sin \phi$ ($|\Delta|>w_0\sin \phi$) such that initial  and final states are respectively gapped and gapless (initial  and final states  become both gapped). The DQPT  occurs for $\mu_i<-2w_0 \cos \phi$ to $-2w_0 \cos \phi<\mu_f<2w_0 \cos \phi$ with $|\Delta|<w_0\sin \phi$ when only $ {h}_{k,f}$ encloses the origin while $ {h}_{k,i}$ and $ {h}_{k,f}$ both have the same chirality (see Fig.~\ref{fig:2} (a) and Fig.~\ref{fig:4}).  The Fisher zeros exhibit 
single crossing on the imaginary axis  ${\rm Re}[z_{n,k}]=0$ such that $t_{c,n}= -iz_{n,k_c} $ (see Fig.~\ref{fig:4} (a)). One can clearly find critical momentum $k_c=1.53$, obtained from Eq.~(\ref{eq:k_c}), across which the  geometric phase $\phi_{k}^{G}(t)$ changes abruptly its sign 
(see Fig.~\ref{fig:4} (c)). The time profile of the winding number exhibits sharp jumps between two quantized plateau exactly at the corresponding critical time $t_{c,n}=\left(n+\frac{1}{2}\right) \pi/h_{k_c,f}$ (see Fig.~\ref{fig:4} (d)). Interestingly, for $-2w_0 \cos \phi<\mu_{i,f}<2w_0 \cos \phi$ with $|\Delta|>w_0\sin \phi$, $h_{k,i}$ and $h_{k,f}$ both encircle the origin and share identical chirality (see Fig.~\ref{fig:2} (b)). We find no DQPT there as the Fisher zeros do not cross the imaginary time axis (see Figs.~\ref{fig:4_2} (a) and (b)).


The encasement of origin for the contour profiles associated with underlying Hamiltonians can be considered to be the necessary condition for DQPT; on the other hand, the opposite chirality of their profiles can be regarded as the 
sufficient condition. We can define a DQPT marker as $\eta= \eta_i-\eta_f$ with $\eta_{i,f}= \pm q_{i,f}$. Here $q_{l}=1$ ($q_{l}=0$) refers to the situation when 
the contour of Hamiltonian ${\cal H}_l$ does (does not) enclose the origin and $+$ ($-$) sign denotes the positive (negative) chirality of the corresponding contour. We investigate various instances where we find that DQPT persists irrespective of the gapless region as long as the system is quenched across or to the QCP $\mu=\pm 2 w_0 \cos \phi$. The DQPT is also observed when the system is quenched from $\Delta_i$ to $\Delta_f$ across the line $\Delta=0$ keeping $\mu$ fixed. All the above observations
are correctly explained by the analysis  $\eta= \eta_i-\eta_f \ne 0 $ whenever the DQPT is observed. According to the above analysis,  the DQPT is accompanied by the crossing of the QCP, separating the two different topological phases or a topological phase from a non-topological phase. This analysis is found to hold true as long as at least one of the Hamiltonian ${\cal H}_{i}$ and ${\cal H}_{f}$ lies outside the gapless phase.


\begin{figure}[H]\includegraphics[trim=1cm 5.3cm 0.7cm 5.5cm, clip=true, height=!,width=1\columnwidth]{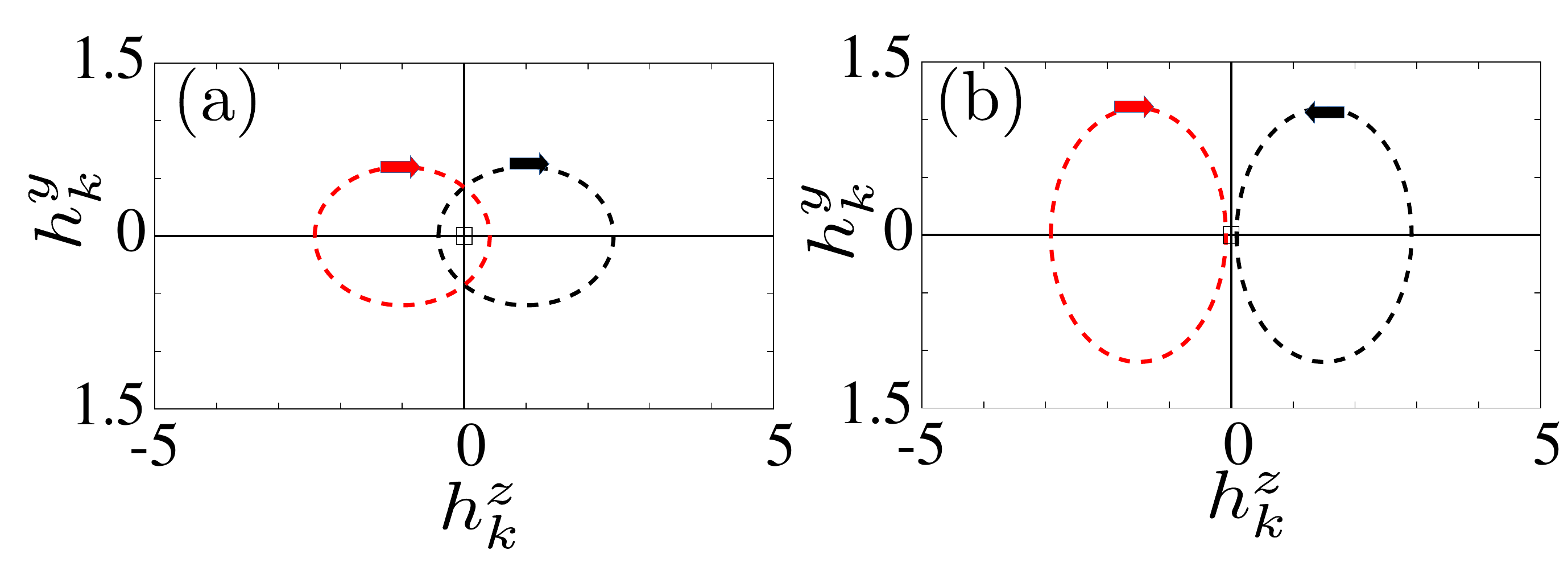}
     \caption{We repeat Fig.~\ref{fig:2} (a) and (b) for            $(\Delta_i,\Delta_f,\mu_i,\mu_f)=(0.3,0.3,-1.0,1.0)$ and $(-0.55,0.55,-1.5,-1.5)$ in (a) and (b), respectively, following $\mu$- and $\Delta$-quench without crossing the QCP $\mu=\pm 2 w_0 \cos \phi$ inside the gapless phase. The initial and final contours both include [exclude] origin  in $h^y_k$-$h^z_k$ plane in (a) [(b)] resulting in no change in topology.    }\label{fig:5}
\end{figure}

\begin{figure}[H]\includegraphics[trim=0.1cm 0.9cm 0.4cm 0.4cm, clip=true, height=!,width=1\columnwidth]{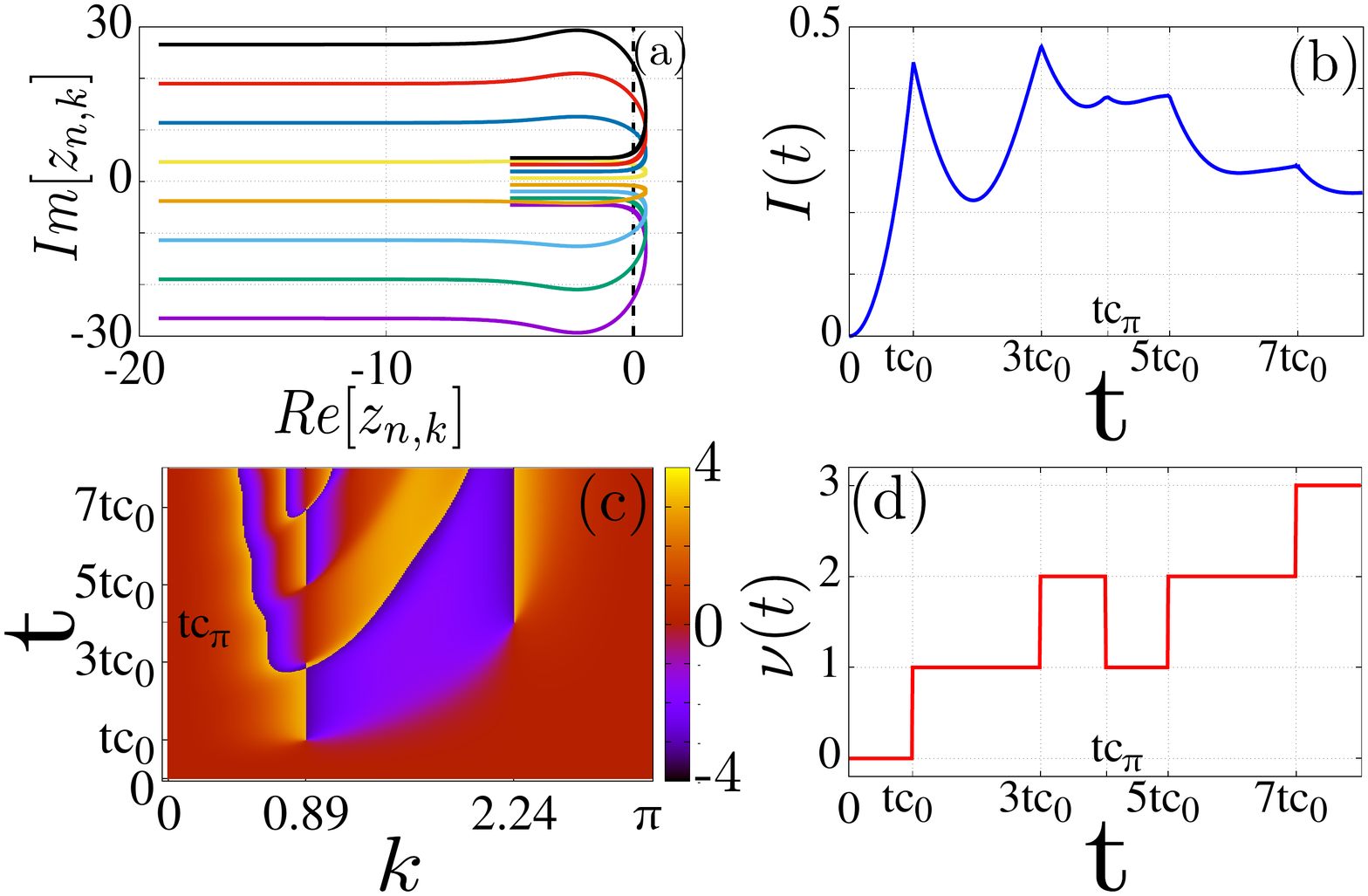}
	\caption{We repeat Fig.~\ref{fig:4} for the case discussed in Fig.~\ref{fig:5} (a). 
	(a) The lines of Fisher zeroes $z_{n,k}$ with $n=-4$ (violet), $\cdots, 4$ (black) cross twice the imaginary axis suggesting the occurrence of DQPT. (b) We find critical times  $t=t_{c0},3t_{c0},t_{c\pi},5t_{c0},7t_{c0}$ at which $I(t)$ diverges with   $t_{c0} \approx 0.81$ and $t_{c \pi} \approx 3.27$. (c) The geometric phase exhibit discontinuous profile at critical momenta $k_{c0}\approx 0.89$ and $k_{c \pi} \approx 2.24$ for the above critical times. (d) The non-monotonic jumps in winding number $\nu(t)$ suggests the existence of two critical time-scales $t_{c0}$ and $t_{c\pi}$ consistent with the rate function. We find qualitatively similar feature for the case of Fig.~\ref{fig:5} (b). 
	   }\label{fig:6}
\end{figure}

We focus on situations  $|\mu|>2w_0\cos \phi$ ($|\mu|<2w_0\cos \phi$) to investigate the emergence of DQPT inside the gapless phase, separating the topological and non-topological phases (two topological phases). According to the above analysis, we find $\eta_i=\eta_f=+1$ ($\eta_i=\eta_f=0$) for $\mu_i=-1 \to \mu_f=1$, $\Delta=0.3$ ($\Delta_i=-0.55 \to \Delta_f=0.55$, $\mu=-1.5$) referring to the fact that DQPT is not expected to occur (see Figs.~\ref{fig:5} (a) and (b)). By contrast, the rate function displays non-analytic behavior as the Fisher zeros cross the ${\rm Re}[z_{n,k}]=0$-axis (see Figs.~\ref{fig:6} (a) and (b)). Importantly, we find two  critical momentum, derived from Eq.~(\ref{eq:k_c}), as follows 
\begin{equation}
\frac{4 \Delta_{i} \Delta_{f}  \sin^{2} k_{c}}{(\mu_{i}+ 2 \cos \phi \cos k_{c} ) (\mu_{f}+ 2 \cos \phi \cos k_{c})} = -1
\label{eq:our_kc}
\end{equation}
yielding $k_{c0} \approx 0.89$, $t_{c0}\approx 0.81$ and $k_{c\pi} \approx 2.24$, $t_{c\pi}\approx 3.27$. The profile of the geometric phase  can be  explained by the above analysis (see Figs.~\ref{fig:6} (c)).
Remarkably, the winding number shows non-monotonic behavior is due to the re-entrant nature of the Fisher zeros where $z_{n,k}$ intersects the imaginary axis twice. The winding number rises (falls) by unity at certain integer multiples of $t_{c0}$ ($t_{c\pi}$) [see Figs.~\ref{fig:6} (d)]. Notice that even though the contour analysis fails to indicate the occurrence of DQPTs that 
is caused by the vanishing $\ln (\tan^2 \phi_k)$-term with $|\sin\phi_k|=|\cos\phi_k|=1/\sqrt{2}$. 
This signals  the infinite
temperature state when both levels of the two-level system are
equally populated. This is what is exactly observed in DQPT, mediated by slow quenching  across the QCP for transverse Ising model, with Landau-Zener transition probability being half \cite{SS}.

Combining all these, we find that contour analysis fails inside the green gapless region when both ${\cal H}_{k,i}$ and ${\cal H}_{k,f}$ reside in the above region. By contrast, inside the gapped region, the contour analysis can successfully predict the occurrences of  DQPTs al least when $\Delta, \mu$, and  $w_0$ are comparable. Therefore, the effect of the gapless region is visible in the DQPTs. However, the DQPT is unable capture any details of the underlying phase. It is noteworthy that as   long as Eq.~(\ref{eq:k_c}) is satisfied the DQPT is bound to happen irrespective of the specific details of the phase. Interestingly, the re-entrant behavior of Fisher zeros lead to non-monotonic behavior of winding number which might be another hallmark signature of the gapless phase. In general,  such re-entrant behavior of   Fisher zeros
is not expected to observe inside the gapped phase.


\subsection{Hybrid case:}
\label{results2}

\begin{figure}[H] 
	\includegraphics[trim=0.9cm 0.55cm 0.2cm 0.5cm, clip=true, height=!,width=1\columnwidth]{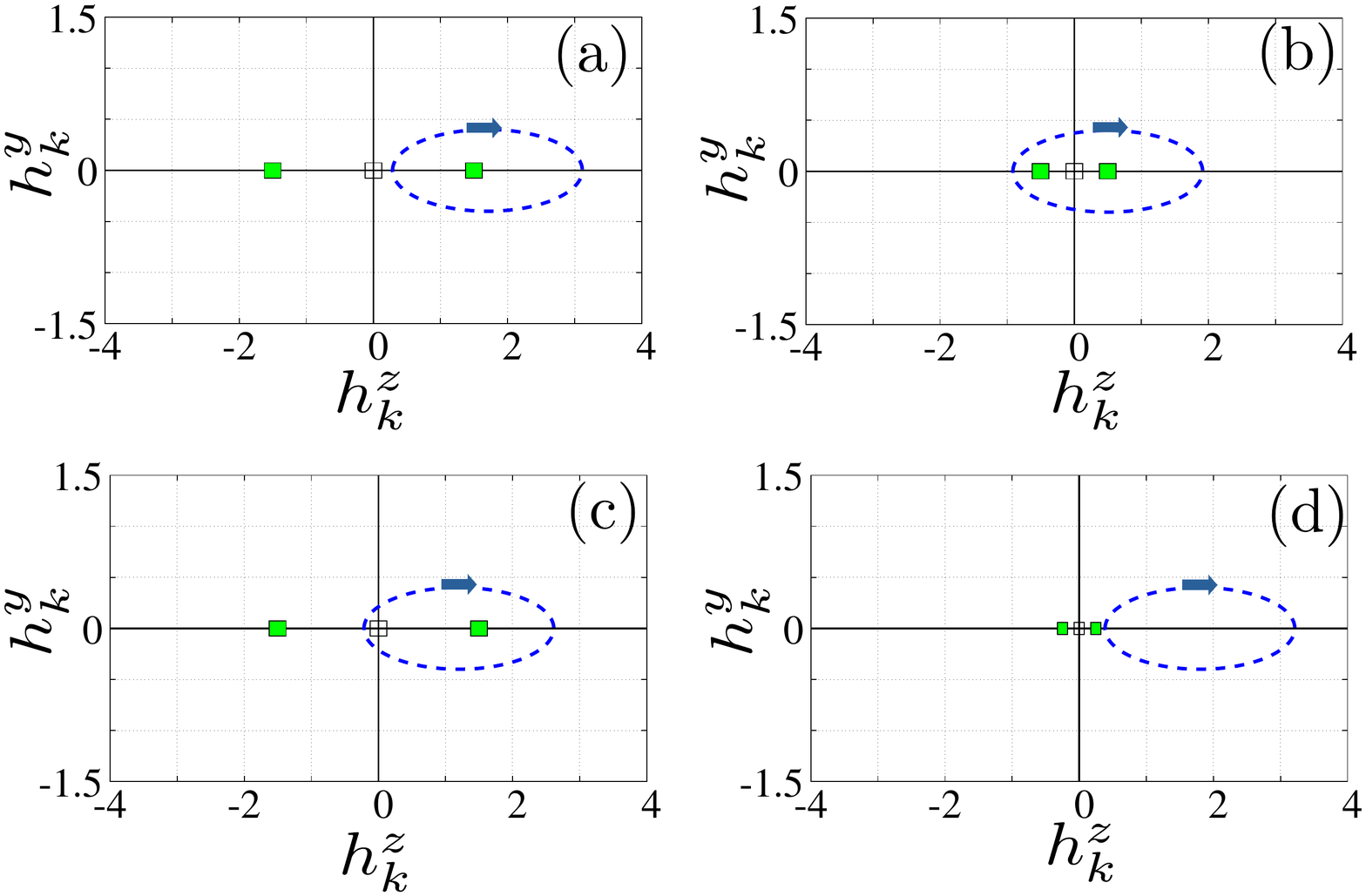}
	\caption{The parametric profiles of $h^y_{k}$ and $h^z_{k}$ for final Hamiltonian ${\mathcal H}_{k,f}={\mathcal H}_{k,i}+i\gamma\sigma_y/2$ with 
	$(\Delta_i,\mu_i,\gamma)=(0.2,-1.7,3.0)$, $(0.2,-0.5,1.0)$, $(0.2,-1.2,3)$ and $(0.2,-1.8,0.5)$ in (a), (b),(c) and (d), respectively. The solid green 
    [empty] rectangles at $(h^y_k,h^z_k)=(0,\pm \gamma/2)$ [$(0,0)$] represent the EPs [origin] for ${\mathcal H}_{f,k}$ [${\mathcal H}_{i,k}$].     }\label{fig:8}
\end{figure}

We now explore the situation where the initial and final Hamiltonians are respectively Hermitian and non-Hermitian: ${\cal H}_{k,f}={\cal H}_{k,i}+i\gamma\sigma_y/2$ keeping all the remaining parameters unaltered.   Note that the initial and final contour are same except the fact that the final contour can only enclose EPs while the initial contour can only encircle the origin i.e, QCP.
Note that for $-2 w_0 \cos \phi + \gamma/2<\mu<2 w_0 \cos \phi - \gamma/2$, ${\cal H}_{k,f}$ encloses  two EPs at $(h^y_{k,f},h^z_{k,f})=(0,\pm \gamma/2)$ simultaneously irrespective of the horizontal gapless phase
boundaries $\Delta =\pm \sqrt{w^2_{0} \sin^2 \phi + \gamma^2/16}$.
A single EP,
appearing at $(0,- \gamma/2)$  [$(0, \gamma/2)$], is enclosed by the Hamiltonian ${\cal H}_{k,f}$ for  $-2 w_0 \cos \phi - \gamma/2<\mu<-2 w_0 \cos \phi + \gamma/2$ [$2 w_0 \cos \phi - \gamma/2<\mu<2 w_0 \cos \phi + \gamma/2$]. On the other hand, the
Hamiltonian ${\cal H}_{k,f}$ does not enclose 
any of the EPs for $|\mu|>|2 w_0 \cos \phi + \gamma/2|$.   
We consider $\eta_f=\pm \sum_{l} q_l$  where  $l$ and $\pm$ denote the number of EPs  and chirality for the contour of ${\cal H}_{k,f}$, respectively; $q_l=1(0)$ for one (no) EP inside the contour.  On the other hand,  $\eta_i= \pm 1 (0)$ when ${\cal H}_{k,i}$ includes (excludes) origin as already described previously.
Therefore, DQPT is expected to occur as long as the final contour encloses EPs even if initial  contour  does not enclose the origin. The chirality of the contour remains unaltered for  ${\cal H}_{k,i}$ and ${\cal H}_{k,f}$. 
We find $\eta=\eta_i-\eta_f=-1$ for $|\mu|>2w_0\cos \phi$ [$|\mu|<2w_0\cos \phi$] with $|\Delta|< w_0 \sin \phi$ as evident from Fig.~\ref{fig:8} (a)  [(b)].

\begin{figure}[H]\includegraphics[trim=0.1cm 0.9cm 0.4cm 0.4cm, clip=true, height=!,width=1\columnwidth]{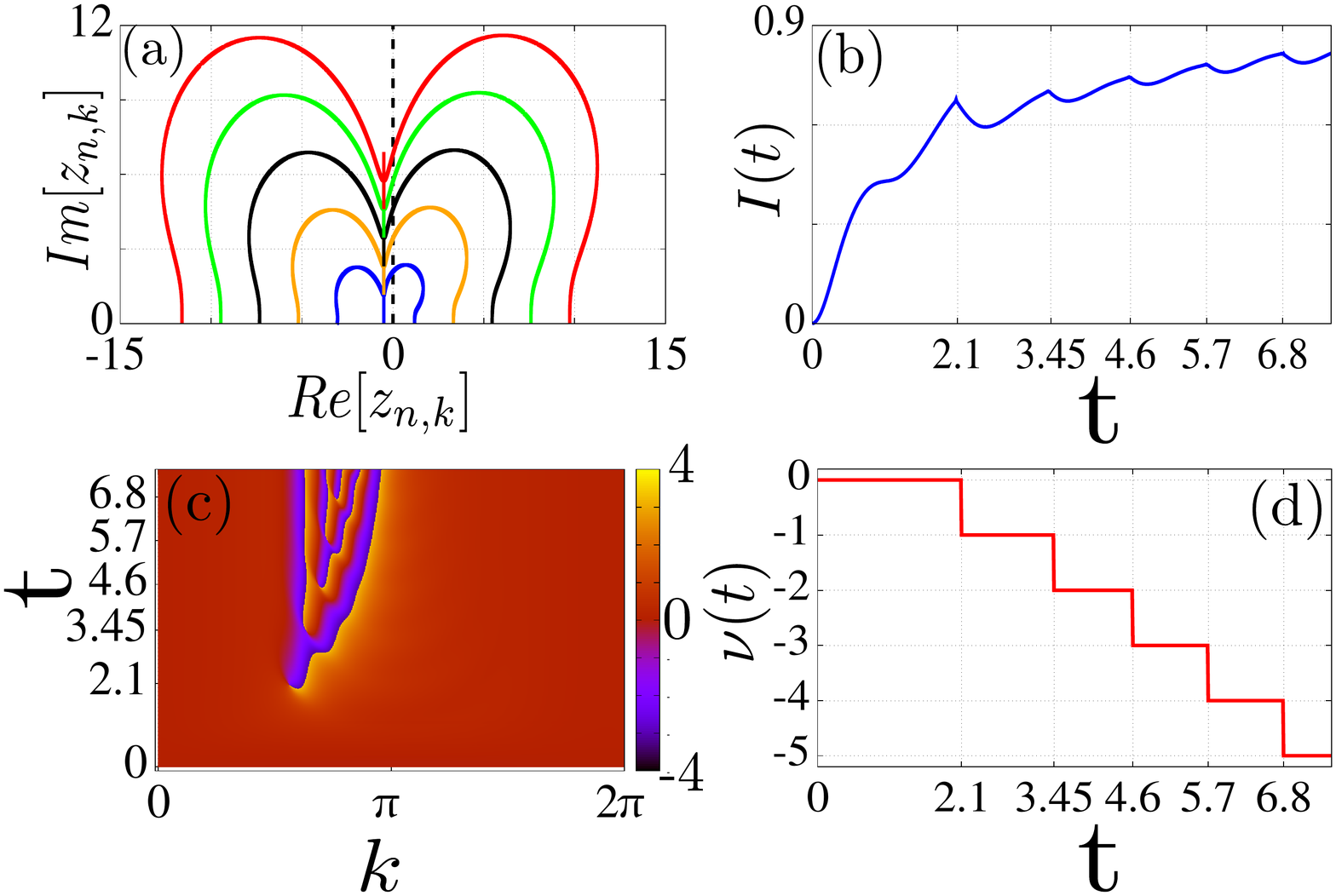}
	\caption{We investigate the DQPTs for the case discussed in Fig.~\ref{fig:8} (a).  
	(a) The  Fisher zeroes $z_{n,k}$ for $n=1$ (blue), $\cdots, n=5$ (red) cross the imaginary axis just once.
	(b) The rate function $I(t)$ diverges at $t \approx 2.1,3.45,4.6,5.7,6.8,\cdots \ne m t_c$ with a fixed value of $t_c$. (c) The geometric phase $\phi_{k}^{G}(t)$ exhibits clear discontinuity  at  several  $k_c$'s yielding above time instants. (d) The winding number $\nu(t)$ decreases monotonically by unit jump at the above time instants as time increases. 
	}\label{fig:9}
\end{figure}

\begin{figure}[H]\includegraphics[trim=0.1cm 0.9cm 0.4cm 0.4cm, clip=true, height=!,width=1\columnwidth]{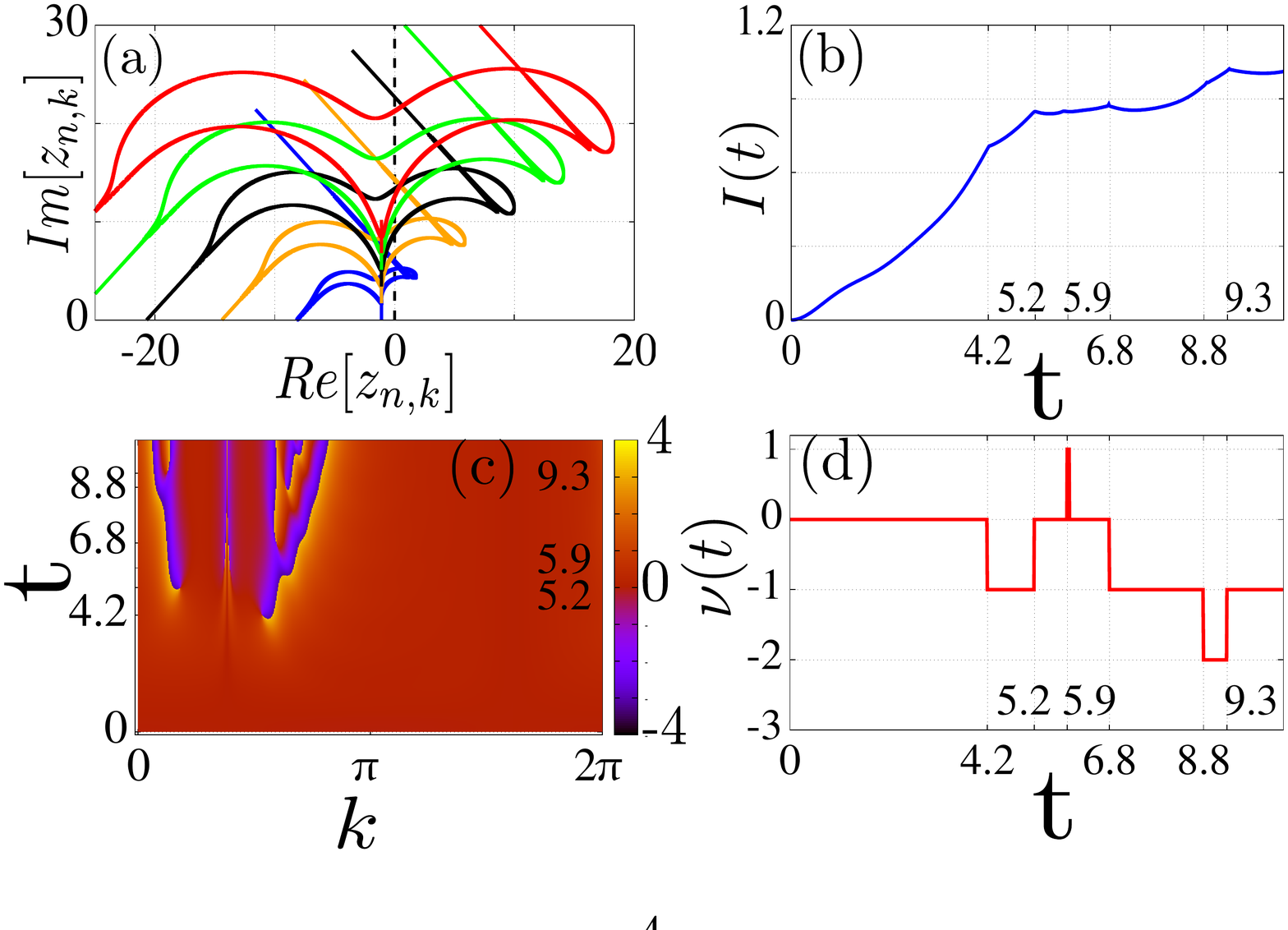}
	\caption{We investigate the DQPTs for the case discussed in Fig.~\ref{fig:8} (b).
	(a) The Fisher zeroes $z_{n,k}$ with $n=1$ (blue), $\cdots,  n=5$ (red) cross real axis four times out of which two  distantly (closely) spaced crossing are caused by two EPs (origin) corresponding to only final Hamiltonian (both final and initial Hamiltonians). (b) The rate function $I(t)$ shows non-analytic behavior at $t \approx 4.2,5.2,5.9,6.8,8.8,9.3, \cdots$. (c) The geometric phase exhibits  three distinct regions where the discontinuties are observed; two of them are corresponding to two EPs and remaining one is for the origin. (d)  The decrease and increase of winding number is related to the EPs while the spike-like behavior  at $t \approx 5.9$ is the signature of the origin. 
	}\label{fig:10}
\end{figure}

A single EP plays the role of origin, as observed for the Hermitian case,  leading to a single crossing of the imaginary axis for Fisher zeros  (see Figs.~\ref{fig:9} (a)). The critical momentum and time can be estimated using the Eqs. (\ref{nh_kc}) and (\ref{nh_tc}), that are clearly depicted in the geometric phase (see Figs.~\ref{fig:9} (c)). The 
non-analyticities in rate function and monotonic jumps in winding number are clearly observed (see Figs.~\ref{fig:9} (b) and (d)). Interestingly, the Fisher zeros cross the imaginary axis four times among which  twice due to two EPs enclosed by ${\cal H}_{k,f}$ and twice for the origin encircled by ${\cal H}_{k,i}$ and ${\cal H}_{k,f}$ both (see Fig.~\ref{fig:10} (a)). The profile of geometric phase, shown in  Fig.~\ref{fig:10} (c),
exhibits distinct signature at certain critical momentum $k_c$ as estimated from  Eq. (\ref{nh_kc}).
The winding number displays positive and negative jumps between two quantized plateau at certain
critical time $t_c$'s, corresponding to two EPs. Interestingly, the kink structure in the winding number is caused by the origin (see Fig.~\ref{fig:10} (d)).  

\begin{figure}[H]\includegraphics[trim=0.1cm 0.7cm 0.4cm 0.4cm, clip=true, height=!,width=1\columnwidth]{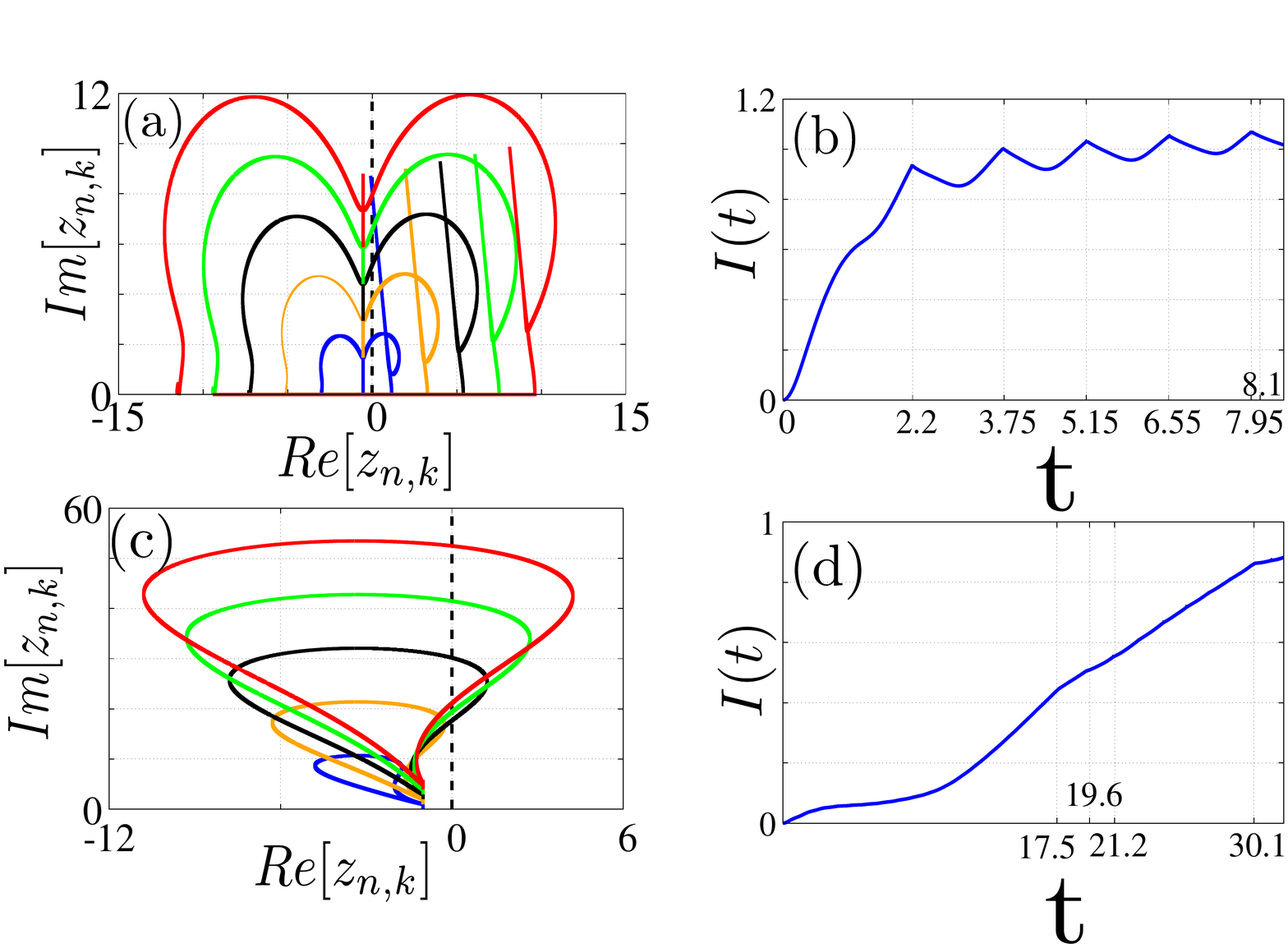}
	\caption{The  Fisher zeros, corresponding to the cases as discussed in Figs.~\ref{fig:8} (c) and (d), are depicted in (a) and (c), respectively. The Fisher zeros cross thrice [twice] the imaginary axis in (a) [(c)]. The corresponding rate functions are displayed in (b) and (d) with $t_c \approx 2.2 , 3.75, 5.15,6.55,7.95,8.1, \cdots$ and $t_c \approx 17.5,19.6,21.2,30.1, \cdots$, respectively.  
	}\label{fig:10_2}
\end{figure}

We find that DQPTs  exist even when the marker $\eta$ vanishes. For example, we adopt a quench with $-2w_0\cos \phi<\mu_i<-2w_0\cos \phi +\gamma/2$ and $\Delta_i<\sqrt{w^2_{0} \sin^2 \phi + \gamma^2/16}$ such that  $\eta_i=\eta_f=1$. This corresponds to a situation when the initial and final contours include origin and one of the EPs, respectively. The Fisher zeros cross the imaginary axis thrice leading to the DQPTs (see Figs.~\ref{fig:10_2} (a) and (b)). Notice that here the final Hamiltonian hosts non-Hermitian gapless phase that was absent for the initial Hermitian Hamiltonian. Therefore, the contour analysis apparently breaks down for the non-Hermitian gapless phase in the present hybrid case. One can further demonstrate an instance with $\eta_i=\eta_f=0$ where DQPT is observed (see Figs.~\ref{fig:10_2} (c) and (d)). The DQPTs are ensured by Eq. (\ref{nh_kc}) even though the contour analysis fails to predict them similar to the previous  Hermitian case. To make the discussion complete, we also investigate the situation with lossy chemical potential ${\cal H}_{k,f}={\cal H}_{k,i}+i\gamma\sigma_z/2$. We find qualitatively similar results as compared to the lossy superconductivity.



\subsection{non-Hermitian case:}
\label{results3}


\begin{figure}[H] 
	\includegraphics[trim=0.9cm 0.55cm 0.2cm 1.5cm, clip=true, height=!,width=1\columnwidth]{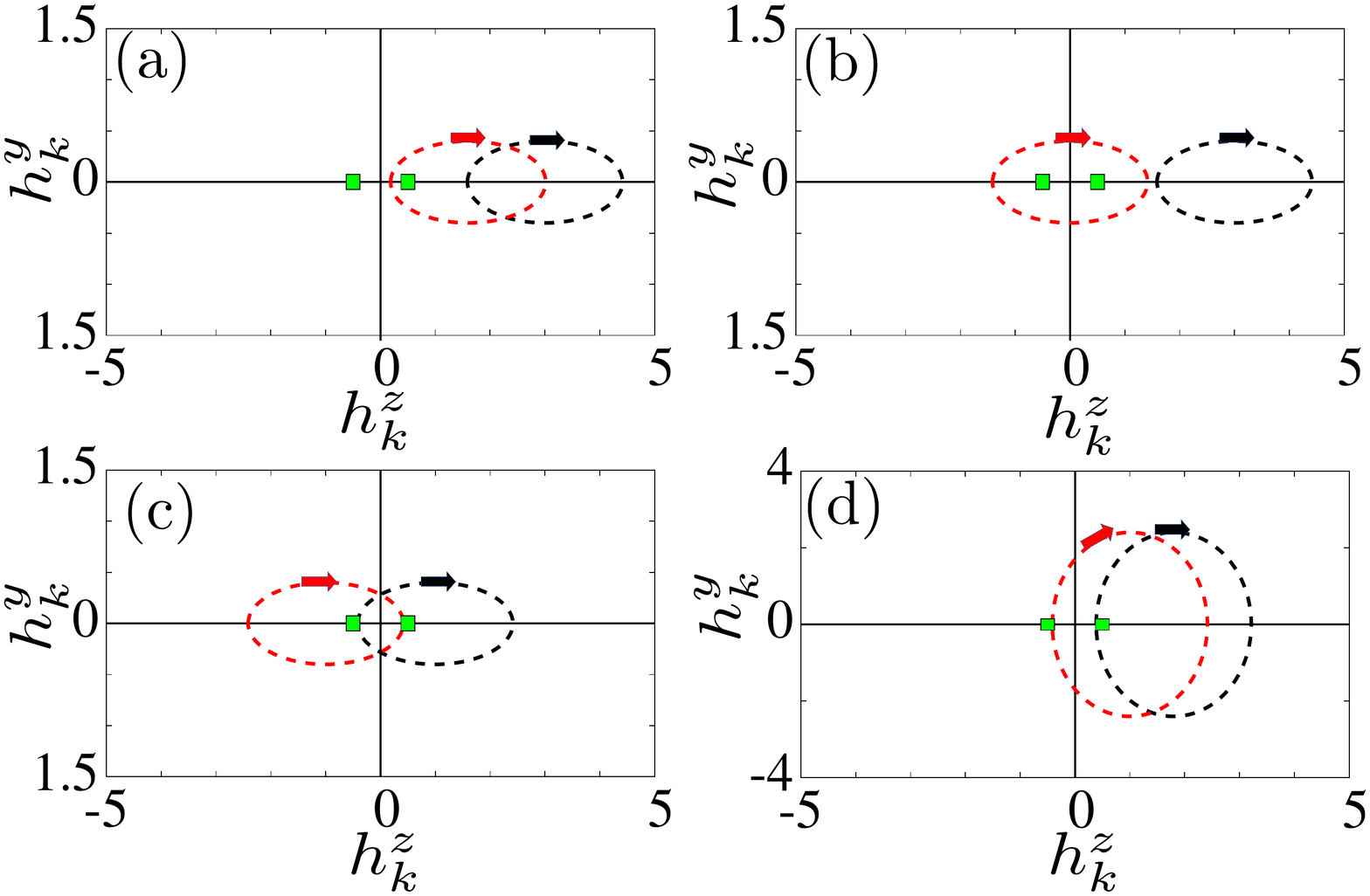}
	\caption{The parametric profiles of $h^y_k$ and $h^z_k$ for  ${\cal H}_{k,i}(0,\gamma,\pi/4)$ (black dashed line) and ${\cal H}_{k,f}(0,\gamma,\pi/4)$ (red dashed line) with $\gamma=1.0$. We consider	$(\Delta_i,\Delta_f,\mu_i,\mu_f)=(0.2,0.2,-3.0,-1.6)$, $(0.2,0.2,-3.0,0.0)$, $(0.2,0.2,-1.0,1.0)$ and $(1.2,1.2,-1.8,-1.0)$ for quenching across one non-Hermitian phase boundary $\mu=-2 w_0 \cos \phi - \gamma/2$ in (a), two non-Hermitian phase boundaries $\mu=-2 w_0 \cos \phi - \gamma/2$ and $\mu=-2 w_0 \cos \phi + \gamma/2$ in (b), in between $\mu=-2 w_0 \cos \phi + \gamma/2$ and $\mu=+2 w_0 \cos \phi - \gamma/2$ in (c), and 
	inside the non-Hermitian phase boundaries $-2 w_0 \cos \phi - \gamma/2<\mu < -2 w_0 \cos \phi + \gamma/2$ in (d), respectively.  The initial black contour encloses no EPs in (a) and (b) while one EP at $(0,\gamma/2)$ in (c) and (d). The final red contour encloses  one EP at $(0,\gamma/2)$ for (a) and (d), $(0,-\gamma/2)$ for (c) and two EPs at  $(0,\pm \gamma/2)$ for (b).
	}\label{fig:11}
\end{figure}


\begin{figure}[H]\includegraphics[trim=0.1cm 0.9cm 0.4cm 0.4cm, clip=true, height=!,width=1\columnwidth]{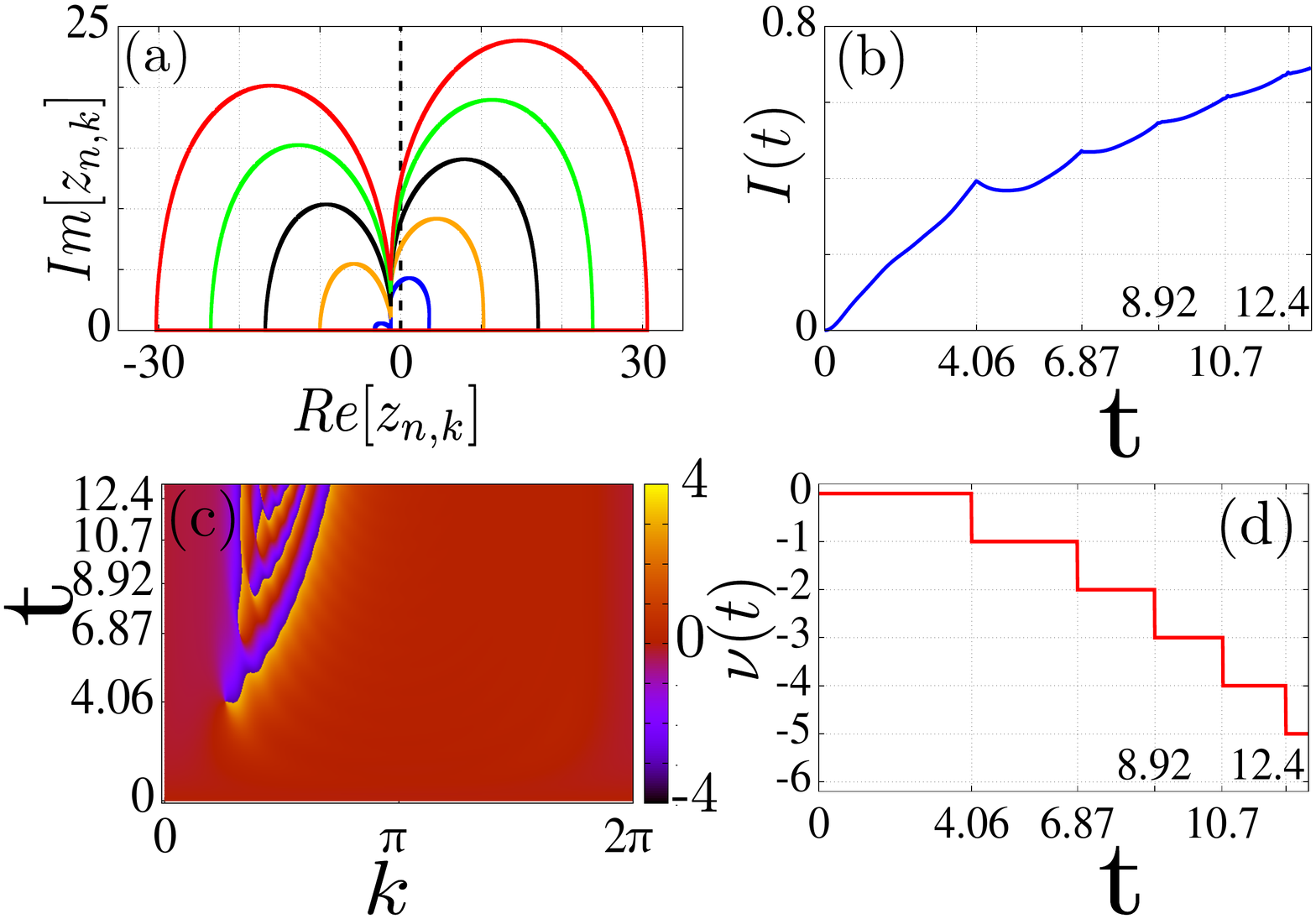}
	\caption{
	We show the Fisher zeros $z_{n,k}$ in (a), rate function $I(t)$ in (b), geometric phase $\phi_{k}^{G}(t)$ in (c),  winding number $\nu(t)$ in (d) for the case as discussed in Fig.~\ref{fig:11}  (a).
	}\label{fig:12}
\end{figure}


\begin{figure}[H]\includegraphics[trim=0.1cm 0.9cm 0.4cm 0.4cm, clip=true, height=!,width=1\columnwidth]{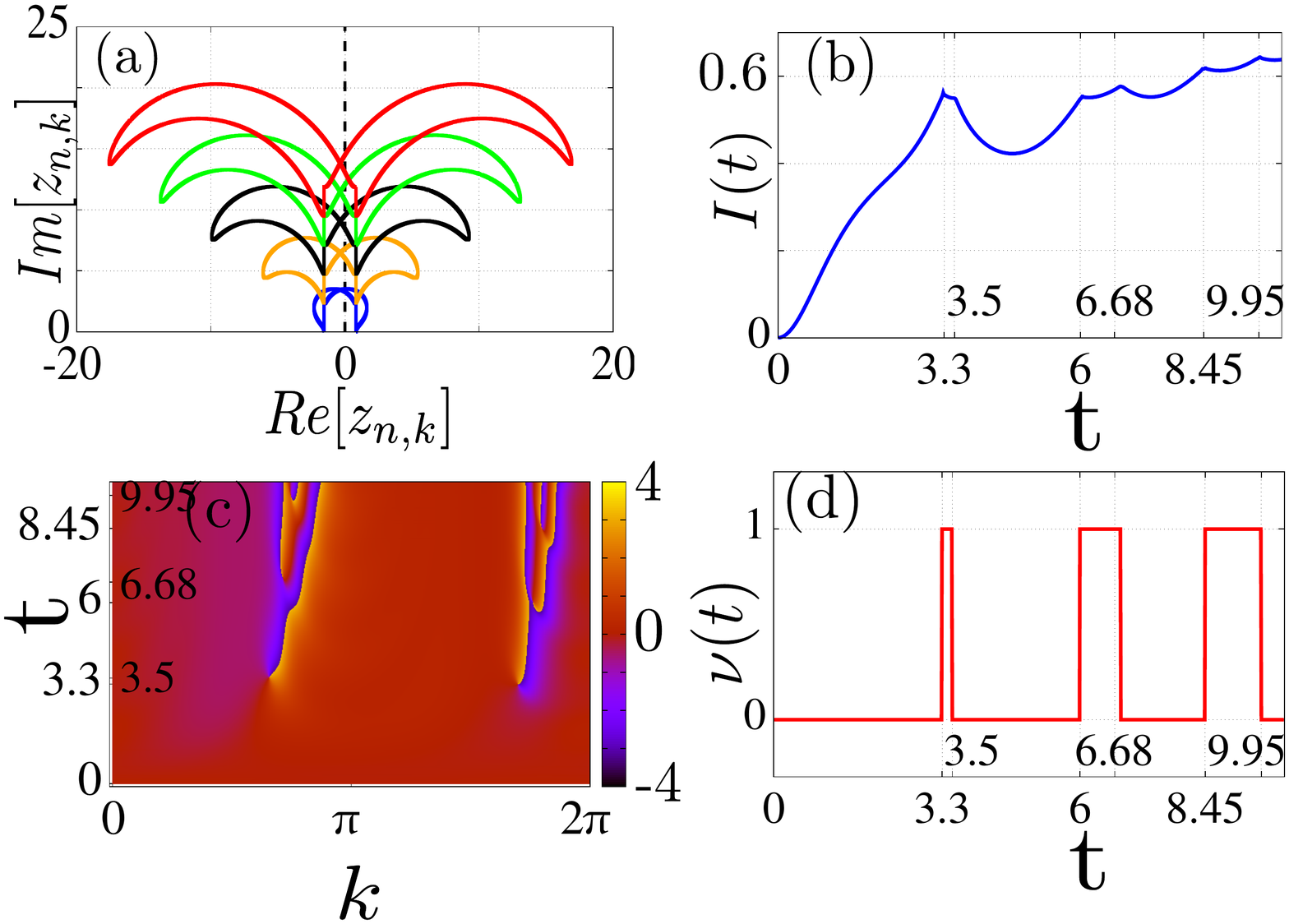}
	\caption{We repeat Fig.~\ref{fig:12} for the case as demonstrated in Fig.~\ref{fig:11}  (b). There exist two critical times 	$t_{c1} \approx 3.3,6.0,8.45,\cdots$ and $t_{c2} \approx 3.5,6.68,9.95, \cdots$ at which $\nu(t)$ increases and decreases, respectively. 
	}\label{fig:13}
\end{figure}

We now consider ${\cal H}_{k,i}(0,\gamma,\phi)$ and ${\cal H}_{k,f}(0,\gamma,\phi)$ both to be non-Hermitian to explore the occurrence of DQPT under various quench path. We consider $\mu$-quench i.e., $\mu_i \to \mu_f$ while $\Delta$ remains fixed i.e, $\Delta_i=\Delta_f$. We only take into account $i \gamma \sigma_y/2$ as the non-Hermiticity while \textsc{$\gamma_1=0$} in Eq.~(\ref{eq:Momentum_Ham}). 
The appearance of the EPs on the non-Hermitian phase diagram is already described in the previous Sec.~\ref{results2}. 
Notice that the quenching from gapped to the gapless phase across the single [double]  non-Hermitian phase boundary [boundaries] at $\mu=-2w_0\cos\phi - \gamma/2$ [$\mu=-2w_0\cos\phi \pm \gamma/2$] are shown in  Fig.~\ref{fig:11} (a) [(b)] where ${\cal H}_{k,f}$ encircles one EP [two EPs] and ${\cal H}_{k,i}$ does not enclose any EPs.
The convention chosen for the contour analysis is the following  $\eta_l=  \pm \sum_l q_l$ with $l$ ($\pm$) denoting the number of EPs inside the contour (chirality for the contour) associated with  ${\cal H}_{k,l}$; $q_l=1(0)$ for one (no) EP inside the contour. Following the contour analysis, $\eta=\eta_i-\eta_f =-1 [-2]$ for Fig.~\ref{fig:11} (a) [(b)] resulting in  DQPTs. 
This yields Fisher zeros to cross once (twice) the imaginary axis and consequently the
winding number exhibit monotonic (non-monotonic) behavior [see Figs.~\ref{fig:12} (a), (d) and Figs.~\ref{fig:13} (a), (d)].
One EP causes the increase in winding number while the
decrease is originated due to the 
other EP.  The profile of geometric phases are also markedly different  for encasing single and double EPs (see Fig.~\ref{fig:12} (c), and Fig.~\ref{fig:13}  (c)). The rate function diverges at certain times $t_c$ for the critical momentum $k_c$, being consistent with Eqs.~(\ref{nh_kc}) and (\ref{nh_tc}), are appropriately captured by geometric phases (see Fig.~\ref{fig:12} (b), and Fig.~\ref{fig:13}  (b)).

\begin{figure}[H]\includegraphics[trim=0.1cm 0.9cm 0.4cm 0.4cm, clip=true, height=!,width=1\columnwidth]{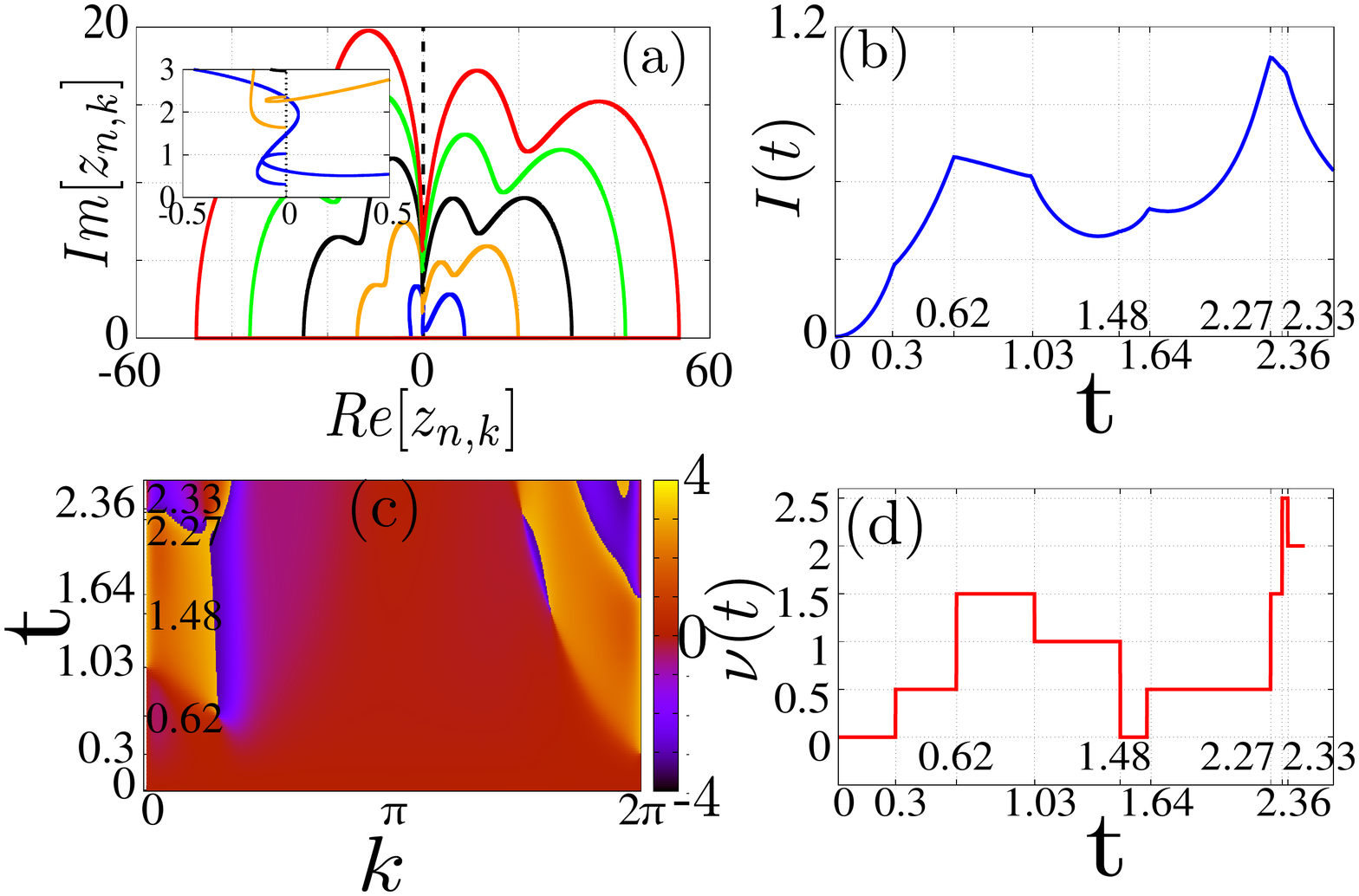}
	\caption{We repeat Fig.~\ref{fig:12} for the case discussed in Fig.~\ref{fig:11}  (c). 
		(a) The Fisher zeroes $z_{n,k}$ with $n=0$ (blue), $\cdots,n=4$ (red) cross imaginary axis continuously three times in addition to two discontinuous crossing. The rate function and geometric phase are depicted in (b) and (c), respectively, where the critical time instants $t_c \approx 0.3,0.62,1.03,1.48,1.64,2.27,2.33,2.36, \cdots$ are mentioned. (d) The winding number $\nu(t)$ jumps by half-integer values ($\pm 0.5$) at  $t_c \approx 0.3,1.03,1.64,2.36,\cdots$ where the Fisher zeros show discontinuities in as shown in the inset of (a). The unit jumps are observed at $t_c \approx 0.62,1.48,2.27,2.33$ corresponding to the continuous crossing of Fisher zeros  over the imaginary axis.        
		}\label{fig:14}
\end{figure}

The contour analysis fails to work i.e., $\eta=0$ for quenching inside the gapless phase bounded by non-Hermitian phase boundaries $-2w_0\cos\phi - \gamma/2< \mu<-2w_0\cos\phi + \gamma/2 $ where the DQPT continues to exist.  Here both the initial and final Hamiltonian enclose the EPs with same chirality (see Fig.~\ref{fig:11} (c)). The Fisher zeros unveils remarkable structure as $z_{n,k}$ crosses imaginary axis in a discontinuous manner (see Fig.~\ref{fig:14} (a)). 
This behavior can be naively understood
by the imaginary values of $k_c=|k_c|e^{i\theta}$ which further leads to multiple values of critical time $t_c$ under the variation of $\theta$. The  critical  momenta $k_c$ and the corresponding time $t_c$ can be obtained from  Eqs.~(\ref{nh_kc}) and (\ref{nh_tc}), respectively. 
Such variation in $\theta$ can be qualitatively understood by the $n$ dependent $k_c$ in Eq.~(\ref{nh_kc}). The profile of the geometric phase is markedly different from the previous cases as we find $k_c\approx 0+\varepsilon $ and $ 2 \pi-\varepsilon $ ($\varepsilon \to 0$) correspond to $t_c\approx 1.03$ and $0.3$
respectively (see Fig.~\ref{fig:14} (c)).
There exist multiple values of time $t_c$ for a given value of $|k_c|$ around which $\phi^G_k(t)$ changes its sign abruptly.
This is intimately related to the jump profile of winding number. Remarkably, we find half-integer jumps associated with the discontinuity in the $z_{n,k}$-profile (see Fig.~\ref{fig:14} (d)). This is strikingly different as compared to all the previous cases where  
the  Fisher zeros, continuously crossing imaginary axis,
always lead to integer jumps.  One can think of that a continuous crossing of Fisher zeros is split into two discontinuous touching of Fisher zeros on the imaginary axis.


\begin{figure}[H]\includegraphics[trim=0.1cm 0.9cm 0.4cm 0.4cm, clip=true, height=!,width=1\columnwidth]{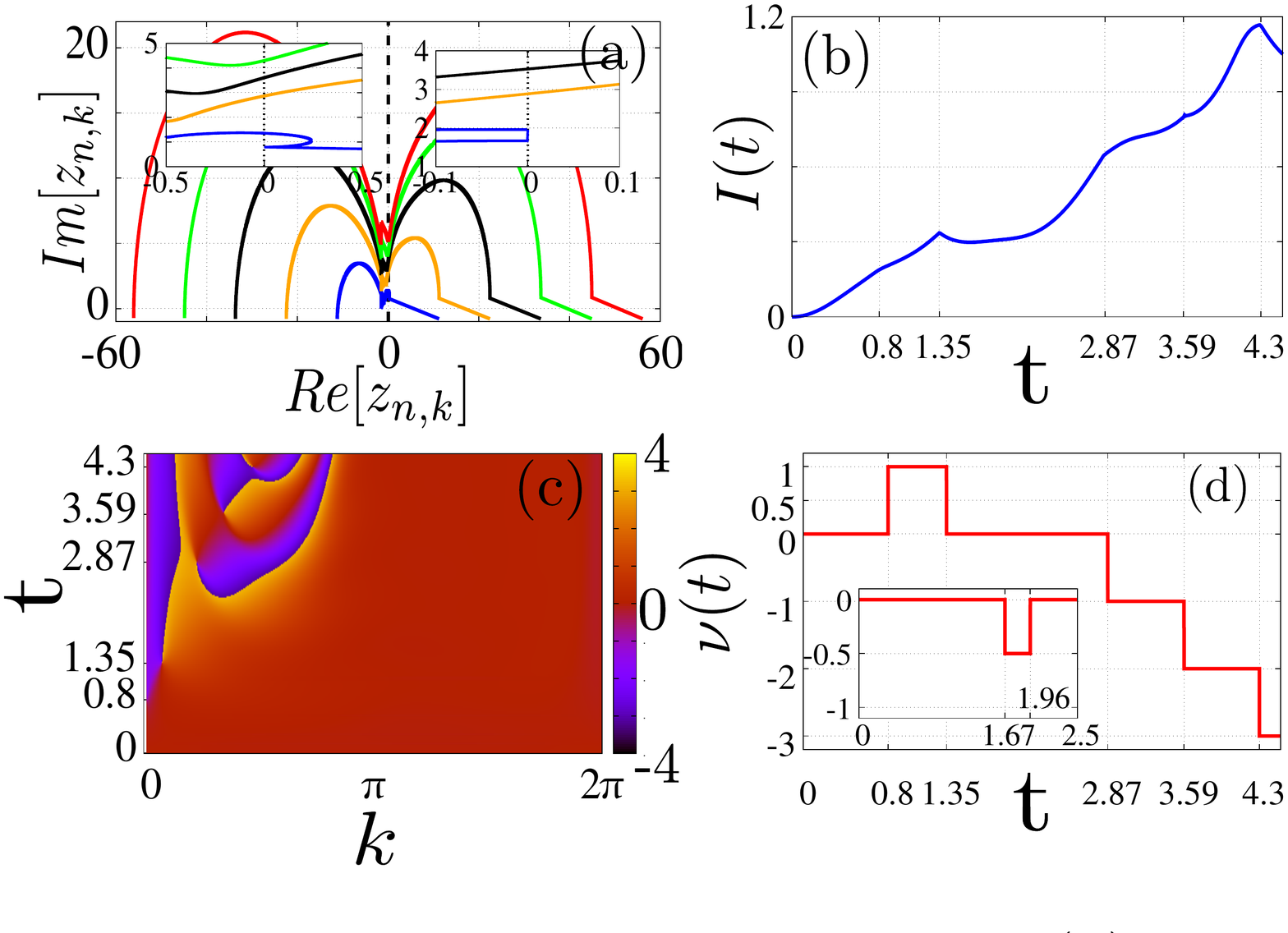}
	\caption{We repeat Fig.~\ref{fig:12} for the case discussed in Fig.~\ref{fig:11}  (d). The Fisher zeros cross the imaginary  axis once continuously while they touch imaginary axis in a quasi-continuous manner as shown in the inset for $n=0$ (blue). The right inset shows the discontinuity in Fisher zeros for $\phi=0$. The rate function and geometric phase are shown in (b) and (c), respectively, where critical time instants $t_c \approx 0.8,1.35,2.87, 3.59, 4.3, \cdots$ are mentioned. (d) The winding number show monotonic decrease (non-monotonic increase) by unit jumps when Fisher zeros continuously cross (quasi-continuously touch) the  imaginary  axis at $t_c \approx 1.35,2.87, \cdots$ [$t_c \approx 0.8$].
	}\label{fig:15}
\end{figure}

We now consider the quench inside the non-Hermitian gapless phase with $-2 w_0\cos \phi - \gamma/2< \mu < -2 w_0\cos \phi + \gamma/2$ that lie above the horizontal gapless region $|\Delta|< \sqrt{(w_0 \sin \phi)^2 +(\gamma/2)^2}$. According to the contour analysis, the EP at $(0,\gamma/2)$ is enclosed by ${\cal H}_{k,i}$ and ${\cal H}_{k,f}$  with the same chirality. The DQPT is thus not expected to occur, however, the Fisher zeros cross imaginary axis leading to non-analytic  signature (jump profile) in rate-function (winding number) [see Figs.~\ref{fig:15} (a), (b), (c) and (d)]. Importantly, half-integer jump is not clearly observed when the Fisher zeros show quasi-continuous profile unlike the previous case. However, the Fisher zeros show clear discontinuity  for $\phi=0$ referring to the fact that
winding number shows half-integer jump (see right inset in Fig.~\ref{fig:15}(a) and inset in Fig.~\ref{fig:15} (d)).  Therefore, the gapless region, originated solely due to the non-Hermiticity, can induce DQPT even though contour analysis fails. The half-integer jumps in winding number is the  hallmark signature for the gapless phases in the non-Hermitian Hamiltonian with lossy superconductivity.

\begin{figure}[H]\includegraphics[trim=0.1cm 0.9cm 0.4cm 0.4cm, clip=true, height=!,width=1\columnwidth]{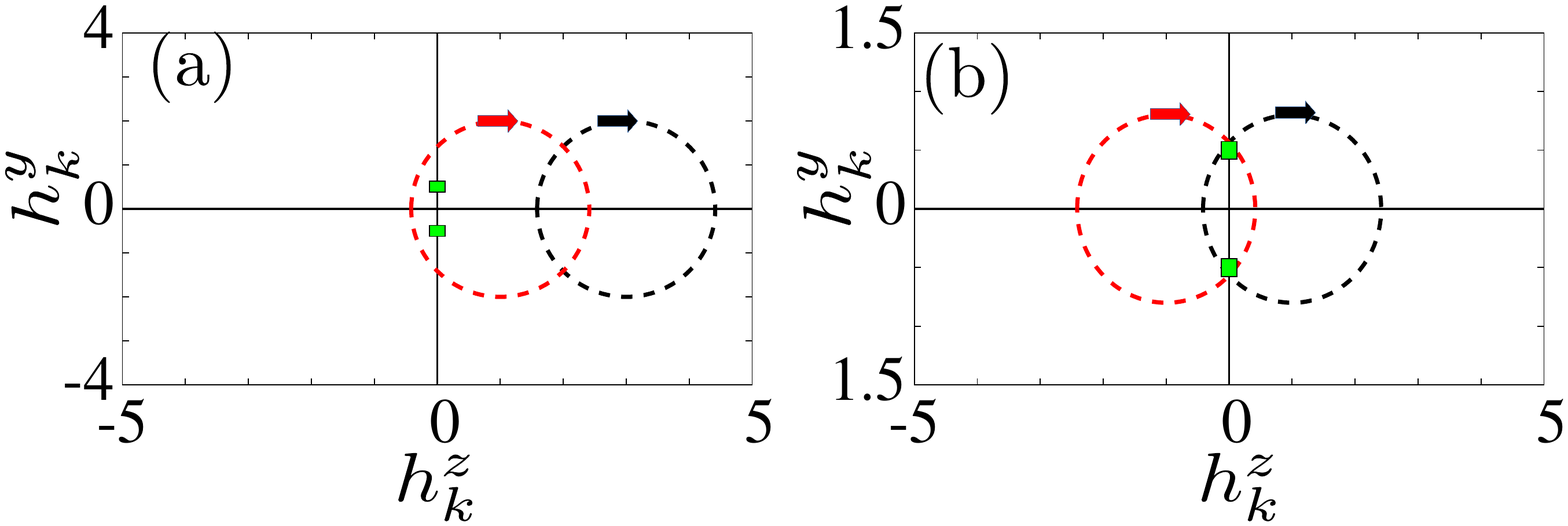}
	\caption{We show the parametric plot for lossy chemical potential case with ${\mathcal H}_{k,i}(\gamma,0,\phi)$ and ${\mathcal H}_{k,f}(\gamma,0,\phi)$ while quenching  across (a) one non-Hermitian phase boundary 
	$\mu=-2w_0 \cos \phi -\gamma/2$ from gapped to gapless phase
    with  $(\Delta_i,\Delta_f,\mu_i,\mu_f)=(1.0,1.0,-3.0,-1.0)$ and (b) 
    two non-Hermitian phase boundaries $\mu=-2w_0 \cos \phi +\gamma/2$ and 
    $\mu=2w_0 \cos \phi -\gamma/2$
    inside the gapless phase with	$(\Delta_i,\Delta_f,\mu_i,\mu_f)=(0.4,0.4,-1.0,1.0)$.  We consider $\gamma=1$. 
	}\label{fig:16}
\end{figure}

\begin{figure}[H]\includegraphics[trim=0.1cm 0.9cm 0.4cm 0.4cm, clip=true, height=!,width=1\columnwidth]{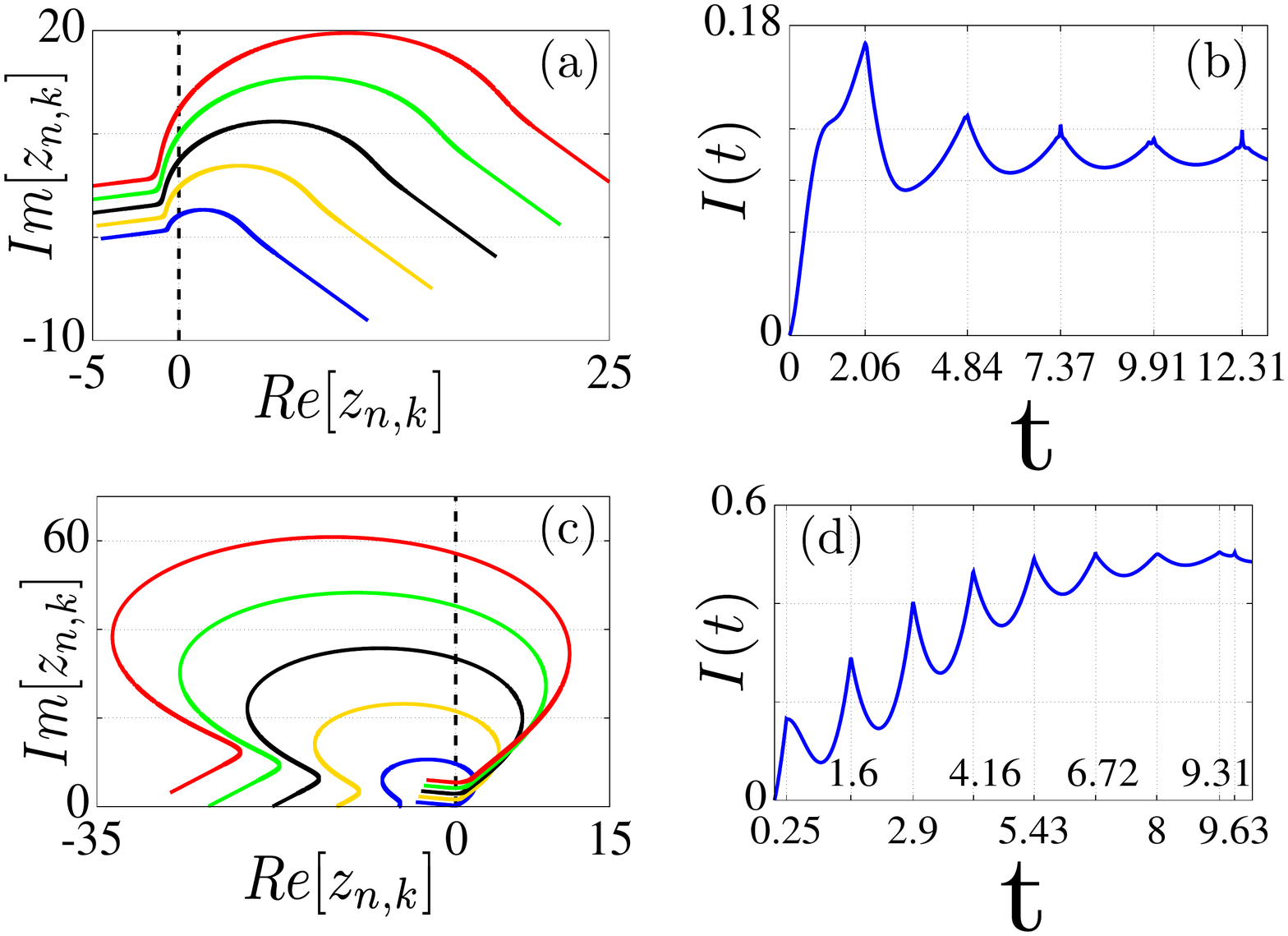}
	\caption{We demonstrate the Fisher zeros  and rate function in (a) [(c)] and (b) [(d)], respectively,  for the case discussed in Fig.~\ref{fig:16} (a) [(b)]. 
	}\label{fig:17}
\end{figure}

Finally, we consider sudden quench from ${\cal H}_{k,i}(\gamma_1,0,\phi)$ to ${\cal H}_{k,f}(\gamma_1,0,\phi)$ with $\gamma=1$. We consider $\mu$-quench i.e., $\mu_i \to \mu_f$ while $\Delta$ remains fixed i.e, $\Delta_i=\Delta_f$. The contour plot in Fig.~\ref{fig:16}
(a) [(b)] shows that only final [both final and initial] Hamiltonian encloses [enclose] EP. As expected from the contour analysis in Fig.~\ref{fig:16}
(a), the quenching across the non-Hermitian critical point $\mu=-2w_0 \cos \phi -\gamma/2$, the DQPT is observed. The inclusion of one EP leads to a single crossing
over the imaginary axis in the Fisher zeros and subsequent occurrence of DQPTs at the corresponding time instants [see Fig.~\ref{fig:17} (a) and (b)]. On the other hand, for  Fig.~\ref{fig:16} (b), the contour analysis fails yielding $\eta=0$ when the system is quenched from one non-Hermitian gapless phase $-2w_0 \cos \phi -\gamma/2< \mu_i < -2w_0 \cos \phi +\gamma/2$ to the other gapless phase $2w_0 \cos \phi -\gamma/2< \mu_f < 2w_0 \cos \phi +\gamma/2$.  However, the inclusion of two distinct EPs separately by initial and final Hamiltonian results in double crossing of Fisher zeros over the imaginary axis. This further ensures the emergence of DQPTs [see Fig.~\ref{fig:17} (c) and (d)] that in turn leads to the non-monotonic profile of winding number.  Importantly, the half-integer jumps are not observed due to continuous profile of Fisher zeros for the non-Hermitian Hamiltonian with lossy chemical potential  $i\gamma \sigma_z/2$ only.  One can naively understand this by the fact that critical momentum $k_c \ne |k_c|e^{i\theta}$. As a result, multiple values of the critical time $t_c$ are not expected to appear for a given value of $|k_c|$. 

Notice that irrespective of the choice of the non-Hermiticity, the the width of gapless phase, bounded by  $-2w_0 \cos \phi +\gamma/2< \mu < 2w_0 \cos \phi -\gamma/2$, extends due to non-Hermiticity. We always have two EPs between the above two non-Hermitian phase boundaries irrespective of the gapless phase. 
The contour analysis fails to indicate the DQPTs for $|\Delta|<\sqrt{w_0^2 \sin^2 \phi + \gamma^2/16}$ [$|\Delta|<\sqrt{w_0^2 \sin^2 \phi + \gamma^2/16-\mu^2/4}$] when initial and final Hamiltonian both reside in the gapless region for lossy chemical potential [superconductivity]. On the other hand, for $|\Delta|>\sqrt{w_0^2 \sin^2 \phi + \gamma^2/16}$ and $|\Delta|>\sqrt{w_0^2 \sin^2 \phi + \gamma^2/16-\mu^2/4}$ when initial and final Hamiltonian both reside in the gapped region, the DQPT is not observed for lossy chemical potential and superconductivity, respectively, as predicted by the 
contour analysis. This feature is similar to that for the Hermitian system. What is more interesting in the non-Hermitian cases is that the gapless phases  bounded by $-2w_0 \cos \phi -\gamma/2< \mu < -2w_0 \cos \phi +\gamma/2$ and $2w_0 \cos \phi -\gamma/2< \mu < 2w_0 \cos \phi +\gamma/2$
only appears due to non-Hermiticity irrespective of their specific types. The quenching inside or between these two phases leads to DQPTs breaking the notion of  the contour analysis. The most interesting aspect is that lossy superconductivity can only result in half-integer jumps in the winding number.

\section{Conclusion}
\label{conclusion}

\begin{table*}[ht]
	\centering 
	\begin{tabular}{| p{2.1cm}|p{2.5cm}|p{1.6cm}|p{3.3cm}|p{1.5cm}|p{1.5cm}|}
		\hline 
		Case  &\hspace{0.0 cm}  Quenching path &\hspace{0.0 cm} Contour analysis & Fisher zeros profile &  Winding number &Example \\ [2ex] 
		\hline 
		Hermitian &\hspace{0.0 cm} $\mu$ quench&\hspace{0.0 cm} valid  & crossing imaginary axis once &$+1$ or $-1$& Fig.\ref{fig:4}\\
		Hermitian &\hspace{0.0 cm} Within gapless region &\hspace{0.0 cm} invalid & crossing imaginary axis twice & $\pm 1$&Fig.\ref{fig:6}\\
		Hybrid($\gamma_2$) &\hspace{0.0 cm} Most cases&\hspace{0.0 cm} valid & crossing imaginary axis once or twice or four times with spike  & $\pm 1$, unit spike & Fig.\ref{fig:9},\ref{fig:10}\\
		Hybrid($\gamma_2$) &\hspace{0.0 cm} Within gapless region&\hspace{0.0 cm} invalid &crossing imaginary axis once or twice or thice with spike  & $\pm 1$,unit spike&Fig.\ref{fig:10_2}\\
		Non-Hermitian($\gamma_2$) &\hspace{0.0 cm} $\mu$ quench&\hspace{0.0 cm} valid &crossing imaginary axis once or twice & $\pm 1$& Fig. \ref{fig:12},\ref{fig:13} \\
		Non-Hermitian($\gamma_2$) &\hspace{0.0 cm} Within gapless region&\hspace{0.0 cm} invalid &discontinuous at imaginary axis  & $\pm 1,\pm \frac{1}{2}$& Fig. \ref{fig:14},\ref{fig:15}\\
		Non-Hermitian($\gamma_1$) &\hspace{0.0 cm} $\mu$ quench&\hspace{0.0 cm} valid & crossing imaginary axis once or twice& $+1 $ or $-1$& Fig. \ref{fig:17}\\
		Non-Hermitian($\gamma_1$) &\hspace{0.0 cm} Within gapless region&\hspace{0.0 cm} invalid& crossing imaginary axis once or twice or four times & $\pm 1$& Fig. \ref{fig:17}\\[2ex] 
		\hline 
	\end{tabular}
	\caption{Table demonstrates the global picture of DQPTs in Hermitian Sec.~\ref{results1}, hybrid Sec.~\ref{results3} and non-Hermitian Sec.~\ref{results3} cases. } 
	\label{table:3} 
\end{table*}


We consider $p$-wave superconductor with complex hopping  to investigate the effect of gapless phases on the DQPTs that is defined by the logarithm of LA. The DQPTs are ensured when the Fisher zeros cross the imaginary axis.   
We make use of the gap terms to designate the contours of the initial and final Hamiltonian involved in the LA following the sudden quench. We find that for Hermitian case, the contour analysis is successful to predict the occurrences of DQPT, associated with the vanishing real part of the Fisher zeros for a given channel, inside the gapped phases except 
the gapless phases (see Figs.~\ref{fig:2}, ~\ref{fig:4}, and \ref{fig:5}). Interestingly, re-entrant profile of Fisher zeros in the gapless region further causes the non-monotonic profile of winding number  that is exclusively observed for the above region only (See Fig.~\ref{fig:6}). Having understood the effect of Hermitian gapless phase, we now extend the analysis to the non-Hermitian gapless regions. For the hybrid case where initial and final Hamiltonians are respectively Hermitian  and non-Hermitian, the contour analysis 
fails when the final Hamiltonian resides in the gapless phase (See Fig.~\ref{fig:8}). The Fisher zeros cross the imaginary axis thrice in the above case where the winding number shows integer spike-like jumps in addition to staircase-like behavior (see Fig.~\ref{fig:10_2}). The contour analysis again fails to predict DQPT when the initial and final Hamiltonian both are non-Hermitian (See Fig.~\ref{fig:11}). Remarkably,  for  lossy superconductivity, the Fisher zeros show discontinuous profile resulting in half-integer jumps in the winding number (See Figs.~\ref{fig:14} and ~\ref{fig:15}). Such a distinct feature is exclusively observed for  the non-Hermitian gapless phase with lossy superconductivity while continuous Fisher zeros with integer jumps are noticed  for lossy chemical potential. Our work
thus extends the notion of DQPTs in the context of the non-Hermitian phases.   
In the future, it would be interesting to study the long-range non-interacting models with
different types of non-Hermiticities.  We also note that the determination of phase by analyzing the DQPTs is yet to be fully explored.  Given the experimental advancement on lossy systems 
\cite{Gou20,li2019observation,Zeuner15,weimann2017topologically,Weiwei18,Gao20}, we believe that  our work can become experimentally relevant.

\section{Acknowledgement} 

We acknowledge SAMKHYA (High-Performance Computing Facility provided by the Institute of Physics,
Bhubaneswar) for our numerical computation. We thank to Arjit Saha for useful discussions. DM thanks to Longwen Zhou for helpful comments. DM thanks to Arnob Kumar Ghosh, Sheikh Moonsun Pervez and Subhadip Bisal for technical help.

\bibliography{bibfile}{}

\begin{thebibliography}{80}%
\makeatletter
\providecommand \@ifxundefined [1]{%
 \@ifx{#1\undefined}
}%
\providecommand \@ifnum [1]{%
 \ifnum #1\expandafter \@firstoftwo
 \else \expandafter \@secondoftwo
 \fi
}%
\providecommand \@ifx [1]{%
 \ifx #1\expandafter \@firstoftwo
 \else \expandafter \@secondoftwo
 \fi
}%
\providecommand \natexlab [1]{#1}%
\providecommand \enquote  [1]{``#1''}%
\providecommand \bibnamefont  [1]{#1}%
\providecommand \bibfnamefont [1]{#1}%
\providecommand \citenamefont [1]{#1}%
\providecommand \href@noop [0]{\@secondoftwo}%
\providecommand \href [0]{\begingroup \@sanitize@url \@href}%
\providecommand \@href[1]{\@@startlink{#1}\@@href}%
\providecommand \@@href[1]{\endgroup#1\@@endlink}%
\providecommand \@sanitize@url [0]{\catcode `\\12\catcode `\$12\catcode
  `\&12\catcode `\#12\catcode `\^12\catcode `\_12\catcode `\%12\relax}%
\providecommand \@@startlink[1]{}%
\providecommand \@@endlink[0]{}%
\providecommand \url  [0]{\begingroup\@sanitize@url \@url }%
\providecommand \@url [1]{\endgroup\@href {#1}{\urlprefix }}%
\providecommand \urlprefix  [0]{URL }%
\providecommand \Eprint [0]{\href }%
\providecommand \doibase [0]{http://dx.doi.org/}%
\providecommand \selectlanguage [0]{\@gobble}%
\providecommand \bibinfo  [0]{\@secondoftwo}%
\providecommand \bibfield  [0]{\@secondoftwo}%
\providecommand \translation [1]{[#1]}%
\providecommand \BibitemOpen [0]{}%
\providecommand \bibitemStop [0]{}%
\providecommand \bibitemNoStop [0]{.\EOS\space}%
\providecommand \EOS [0]{\spacefactor3000\relax}%
\providecommand \BibitemShut  [1]{\csname bibitem#1\endcsname}%
\let\auto@bib@innerbib\@empty
\bibitem [{\citenamefont {Fisher}(1967)}]{fisher1967theory}%
  \BibitemOpen
  \bibfield  {author} {\bibinfo {author} {\bibfnamefont {M.~E.}\ \bibnamefont
  {Fisher}},\ }\href@noop {} {\bibfield  {journal} {\bibinfo  {journal}
  {Reports on progress in physics}\ }\textbf {\bibinfo {volume} {30}},\
  \bibinfo {pages} {615} (\bibinfo {year} {1967})}\BibitemShut {NoStop}%
\bibitem [{\citenamefont {Yang}\ and\ \citenamefont {Lee}(1952)}]{LYF1}%
  \BibitemOpen
  \bibfield  {author} {\bibinfo {author} {\bibfnamefont {C.~N.}\ \bibnamefont
  {Yang}}\ and\ \bibinfo {author} {\bibfnamefont {T.~D.}\ \bibnamefont {Lee}},\
  }\href {\doibase 10.1103/PhysRev.87.404} {\bibfield  {journal} {\bibinfo
  {journal} {Phys. Rev.}\ }\textbf {\bibinfo {volume} {87}},\ \bibinfo {pages}
  {404} (\bibinfo {year} {1952})}\BibitemShut {NoStop}%
\bibitem [{\citenamefont {Lee}\ and\ \citenamefont {Yang}(1952)}]{LYF2}%
  \BibitemOpen
  \bibfield  {author} {\bibinfo {author} {\bibfnamefont {T.~D.}\ \bibnamefont
  {Lee}}\ and\ \bibinfo {author} {\bibfnamefont {C.~N.}\ \bibnamefont {Yang}},\
  }\href {\doibase 10.1103/PhysRev.87.410} {\bibfield  {journal} {\bibinfo
  {journal} {Phys. Rev.}\ }\textbf {\bibinfo {volume} {87}},\ \bibinfo {pages}
  {410} (\bibinfo {year} {1952})}\BibitemShut {NoStop}%
\bibitem [{\citenamefont {Heyl}\ \emph {et~al.}(2013)\citenamefont {Heyl},
  \citenamefont {Polkovnikov},\ and\ \citenamefont {Kehrein}}]{heyl13}%
  \BibitemOpen
  \bibfield  {author} {\bibinfo {author} {\bibfnamefont {M.}~\bibnamefont
  {Heyl}}, \bibinfo {author} {\bibfnamefont {A.}~\bibnamefont {Polkovnikov}}, \
  and\ \bibinfo {author} {\bibfnamefont {S.}~\bibnamefont {Kehrein}},\ }\href
  {\doibase 10.1103/PhysRevLett.110.135704} {\bibfield  {journal} {\bibinfo
  {journal} {Phys. Rev. Lett.}\ }\textbf {\bibinfo {volume} {110}},\ \bibinfo
  {pages} {135704} (\bibinfo {year} {2013})}\BibitemShut {NoStop}%
\bibitem [{\citenamefont {Karrasch}\ and\ \citenamefont
  {Schuricht}(2013)}]{PhysRevB.87.195104}%
  \BibitemOpen
  \bibfield  {author} {\bibinfo {author} {\bibfnamefont {C.}~\bibnamefont
  {Karrasch}}\ and\ \bibinfo {author} {\bibfnamefont {D.}~\bibnamefont
  {Schuricht}},\ }\href {\doibase 10.1103/PhysRevB.87.195104} {\bibfield
  {journal} {\bibinfo  {journal} {Phys. Rev. B}\ }\textbf {\bibinfo {volume}
  {87}},\ \bibinfo {pages} {195104} (\bibinfo {year} {2013})}\BibitemShut
  {NoStop}%
\bibitem [{\citenamefont {Kriel}\ \emph {et~al.}(2014)\citenamefont {Kriel},
  \citenamefont {Karrasch},\ and\ \citenamefont
  {Kehrein}}]{PhysRevB.90.125106}%
  \BibitemOpen
  \bibfield  {author} {\bibinfo {author} {\bibfnamefont {J.~N.}\ \bibnamefont
  {Kriel}}, \bibinfo {author} {\bibfnamefont {C.}~\bibnamefont {Karrasch}}, \
  and\ \bibinfo {author} {\bibfnamefont {S.}~\bibnamefont {Kehrein}},\ }\href
  {\doibase 10.1103/PhysRevB.90.125106} {\bibfield  {journal} {\bibinfo
  {journal} {Phys. Rev. B}\ }\textbf {\bibinfo {volume} {90}},\ \bibinfo
  {pages} {125106} (\bibinfo {year} {2014})}\BibitemShut {NoStop}%
\bibitem [{\citenamefont {Canovi}\ \emph {et~al.}(2014)\citenamefont {Canovi},
  \citenamefont {Werner},\ and\ \citenamefont
  {Eckstein}}]{PhysRevLett.113.265702}%
  \BibitemOpen
  \bibfield  {author} {\bibinfo {author} {\bibfnamefont {E.}~\bibnamefont
  {Canovi}}, \bibinfo {author} {\bibfnamefont {P.}~\bibnamefont {Werner}}, \
  and\ \bibinfo {author} {\bibfnamefont {M.}~\bibnamefont {Eckstein}},\ }\href
  {\doibase 10.1103/PhysRevLett.113.265702} {\bibfield  {journal} {\bibinfo
  {journal} {Phys. Rev. Lett.}\ }\textbf {\bibinfo {volume} {113}},\ \bibinfo
  {pages} {265702} (\bibinfo {year} {2014})}\BibitemShut {NoStop}%
\bibitem [{\citenamefont {Heyl}(2015)}]{PhysRevLett.115.140602}%
  \BibitemOpen
  \bibfield  {author} {\bibinfo {author} {\bibfnamefont {M.}~\bibnamefont
  {Heyl}},\ }\href {\doibase 10.1103/PhysRevLett.115.140602} {\bibfield
  {journal} {\bibinfo  {journal} {Phys. Rev. Lett.}\ }\textbf {\bibinfo
  {volume} {115}},\ \bibinfo {pages} {140602} (\bibinfo {year}
  {2015})}\BibitemShut {NoStop}%
\bibitem [{\citenamefont {Heyl}(2018)}]{Heyl_2018}%
  \BibitemOpen
  \bibfield  {author} {\bibinfo {author} {\bibfnamefont {M.}~\bibnamefont
  {Heyl}},\ }\href {\doibase 10.1088/1361-6633/aaaf9a} {\bibfield  {journal}
  {\bibinfo  {journal} {Reports on Progress in Physics}\ }\textbf {\bibinfo
  {volume} {81}},\ \bibinfo {pages} {054001} (\bibinfo {year}
  {2018})}\BibitemShut {NoStop}%
\bibitem [{\citenamefont {Bhattacharya}\ \emph {et~al.}(2017)\citenamefont
  {Bhattacharya}, \citenamefont {Bandyopadhyay},\ and\ \citenamefont
  {Dutta}}]{Bhattacharya17}%
  \BibitemOpen
  \bibfield  {author} {\bibinfo {author} {\bibfnamefont {U.}~\bibnamefont
  {Bhattacharya}}, \bibinfo {author} {\bibfnamefont {S.}~\bibnamefont
  {Bandyopadhyay}}, \ and\ \bibinfo {author} {\bibfnamefont {A.}~\bibnamefont
  {Dutta}},\ }\href {\doibase 10.1103/PhysRevB.96.180303} {\bibfield  {journal}
  {\bibinfo  {journal} {Phys. Rev. B}\ }\textbf {\bibinfo {volume} {96}},\
  \bibinfo {pages} {180303} (\bibinfo {year} {2017})}\BibitemShut {NoStop}%
\bibitem [{\citenamefont {Jafari}\ \emph {et~al.}(2019)\citenamefont {Jafari},
  \citenamefont {Johannesson}, \citenamefont {Langari},\ and\ \citenamefont
  {Martin-Delgado}}]{Jafari19a}%
  \BibitemOpen
  \bibfield  {author} {\bibinfo {author} {\bibfnamefont {R.}~\bibnamefont
  {Jafari}}, \bibinfo {author} {\bibfnamefont {H.}~\bibnamefont {Johannesson}},
  \bibinfo {author} {\bibfnamefont {A.}~\bibnamefont {Langari}}, \ and\
  \bibinfo {author} {\bibfnamefont {M.~A.}\ \bibnamefont {Martin-Delgado}},\
  }\href {\doibase 10.1103/PhysRevB.99.054302} {\bibfield  {journal} {\bibinfo
  {journal} {Phys. Rev. B}\ }\textbf {\bibinfo {volume} {99}},\ \bibinfo
  {pages} {054302} (\bibinfo {year} {2019})}\BibitemShut {NoStop}%
\bibitem [{\citenamefont {Uhrich}\ \emph {et~al.}(2020)\citenamefont {Uhrich},
  \citenamefont {Defenu}, \citenamefont {Jafari},\ and\ \citenamefont
  {Halimeh}}]{Uhrich20}%
  \BibitemOpen
  \bibfield  {author} {\bibinfo {author} {\bibfnamefont {P.}~\bibnamefont
  {Uhrich}}, \bibinfo {author} {\bibfnamefont {N.}~\bibnamefont {Defenu}},
  \bibinfo {author} {\bibfnamefont {R.}~\bibnamefont {Jafari}}, \ and\ \bibinfo
  {author} {\bibfnamefont {J.~C.}\ \bibnamefont {Halimeh}},\ }\href {\doibase
  10.1103/PhysRevB.101.245148} {\bibfield  {journal} {\bibinfo  {journal}
  {Phys. Rev. B}\ }\textbf {\bibinfo {volume} {101}},\ \bibinfo {pages}
  {245148} (\bibinfo {year} {2020})}\BibitemShut {NoStop}%
\bibitem [{\citenamefont {Vajna}\ and\ \citenamefont
  {D\'ora}(2014{\natexlab{a}})}]{Vajna14}%
  \BibitemOpen
  \bibfield  {author} {\bibinfo {author} {\bibfnamefont {S.}~\bibnamefont
  {Vajna}}\ and\ \bibinfo {author} {\bibfnamefont {B.}~\bibnamefont {D\'ora}},\
  }\href {\doibase 10.1103/PhysRevB.89.161105} {\bibfield  {journal} {\bibinfo
  {journal} {Phys. Rev. B}\ }\textbf {\bibinfo {volume} {89}},\ \bibinfo
  {pages} {161105} (\bibinfo {year} {2014}{\natexlab{a}})}\BibitemShut
  {NoStop}%
\bibitem [{\citenamefont {Schmitt}\ and\ \citenamefont
  {Kehrein}(2015{\natexlab{a}})}]{Schmitt15}%
  \BibitemOpen
  \bibfield  {author} {\bibinfo {author} {\bibfnamefont {M.}~\bibnamefont
  {Schmitt}}\ and\ \bibinfo {author} {\bibfnamefont {S.}~\bibnamefont
  {Kehrein}},\ }\href {\doibase 10.1103/PhysRevB.92.075114} {\bibfield
  {journal} {\bibinfo  {journal} {Phys. Rev. B}\ }\textbf {\bibinfo {volume}
  {92}},\ \bibinfo {pages} {075114} (\bibinfo {year}
  {2015}{\natexlab{a}})}\BibitemShut {NoStop}%
\bibitem [{\citenamefont {Halimeh}\ and\ \citenamefont
  {Zauner-Stauber}(2017)}]{Halimeh17}%
  \BibitemOpen
  \bibfield  {author} {\bibinfo {author} {\bibfnamefont {J.~C.}\ \bibnamefont
  {Halimeh}}\ and\ \bibinfo {author} {\bibfnamefont {V.}~\bibnamefont
  {Zauner-Stauber}},\ }\href {\doibase 10.1103/PhysRevB.96.134427} {\bibfield
  {journal} {\bibinfo  {journal} {Phys. Rev. B}\ }\textbf {\bibinfo {volume}
  {96}},\ \bibinfo {pages} {134427} (\bibinfo {year} {2017})}\BibitemShut
  {NoStop}%
\bibitem [{\citenamefont {\ifmmode \check{Z}\else
  \v{Z}\fi{}unkovi\ifmmode~\check{c}\else \v{c}\fi{}}\ \emph
  {et~al.}(2018)\citenamefont {\ifmmode \check{Z}\else
  \v{Z}\fi{}unkovi\ifmmode~\check{c}\else \v{c}\fi{}}, \citenamefont {Heyl},
  \citenamefont {Knap},\ and\ \citenamefont {Silva}}]{Silva18}%
  \BibitemOpen
  \bibfield  {author} {\bibinfo {author} {\bibfnamefont {B.}~\bibnamefont
  {\ifmmode \check{Z}\else \v{Z}\fi{}unkovi\ifmmode~\check{c}\else
  \v{c}\fi{}}}, \bibinfo {author} {\bibfnamefont {M.}~\bibnamefont {Heyl}},
  \bibinfo {author} {\bibfnamefont {M.}~\bibnamefont {Knap}}, \ and\ \bibinfo
  {author} {\bibfnamefont {A.}~\bibnamefont {Silva}},\ }\href {\doibase
  10.1103/PhysRevLett.120.130601} {\bibfield  {journal} {\bibinfo  {journal}
  {Phys. Rev. Lett.}\ }\textbf {\bibinfo {volume} {120}},\ \bibinfo {pages}
  {130601} (\bibinfo {year} {2018})}\BibitemShut {NoStop}%
\bibitem [{\citenamefont {Halimeh}\ \emph {et~al.}(2020)\citenamefont
  {Halimeh}, \citenamefont {Van~Damme}, \citenamefont {Zauner-Stauber},\ and\
  \citenamefont {Vanderstraeten}}]{Halimeh20c}%
  \BibitemOpen
  \bibfield  {author} {\bibinfo {author} {\bibfnamefont {J.~C.}\ \bibnamefont
  {Halimeh}}, \bibinfo {author} {\bibfnamefont {M.}~\bibnamefont {Van~Damme}},
  \bibinfo {author} {\bibfnamefont {V.}~\bibnamefont {Zauner-Stauber}}, \ and\
  \bibinfo {author} {\bibfnamefont {L.}~\bibnamefont {Vanderstraeten}},\ }\href
  {\doibase 10.1103/PhysRevResearch.2.033111} {\bibfield  {journal} {\bibinfo
  {journal} {Phys. Rev. Research}\ }\textbf {\bibinfo {volume} {2}},\ \bibinfo
  {pages} {033111} (\bibinfo {year} {2020})}\BibitemShut {NoStop}%
\bibitem [{\citenamefont {Hashizume}\ \emph {et~al.}(2022)\citenamefont
  {Hashizume}, \citenamefont {McCulloch},\ and\ \citenamefont
  {Halimeh}}]{Hashizume22}%
  \BibitemOpen
  \bibfield  {author} {\bibinfo {author} {\bibfnamefont {T.}~\bibnamefont
  {Hashizume}}, \bibinfo {author} {\bibfnamefont {I.~P.}\ \bibnamefont
  {McCulloch}}, \ and\ \bibinfo {author} {\bibfnamefont {J.~C.}\ \bibnamefont
  {Halimeh}},\ }\href {\doibase 10.1103/PhysRevResearch.4.013250} {\bibfield
  {journal} {\bibinfo  {journal} {Phys. Rev. Research}\ }\textbf {\bibinfo
  {volume} {4}},\ \bibinfo {pages} {013250} (\bibinfo {year}
  {2022})}\BibitemShut {NoStop}%
\bibitem [{\citenamefont {Lang}\ \emph {et~al.}(2018)\citenamefont {Lang},
  \citenamefont {Frank},\ and\ \citenamefont {Halimeh}}]{Lang18}%
  \BibitemOpen
  \bibfield  {author} {\bibinfo {author} {\bibfnamefont {J.}~\bibnamefont
  {Lang}}, \bibinfo {author} {\bibfnamefont {B.}~\bibnamefont {Frank}}, \ and\
  \bibinfo {author} {\bibfnamefont {J.~C.}\ \bibnamefont {Halimeh}},\ }\href
  {\doibase 10.1103/PhysRevB.97.174401} {\bibfield  {journal} {\bibinfo
  {journal} {Phys. Rev. B}\ }\textbf {\bibinfo {volume} {97}},\ \bibinfo
  {pages} {174401} (\bibinfo {year} {2018})}\BibitemShut {NoStop}%
\bibitem [{\citenamefont {Homrighausen}\ \emph {et~al.}(2017)\citenamefont
  {Homrighausen}, \citenamefont {Abeling}, \citenamefont {Zauner-Stauber},\
  and\ \citenamefont {Halimeh}}]{Homrighausen17}%
  \BibitemOpen
  \bibfield  {author} {\bibinfo {author} {\bibfnamefont {I.}~\bibnamefont
  {Homrighausen}}, \bibinfo {author} {\bibfnamefont {N.~O.}\ \bibnamefont
  {Abeling}}, \bibinfo {author} {\bibfnamefont {V.}~\bibnamefont
  {Zauner-Stauber}}, \ and\ \bibinfo {author} {\bibfnamefont {J.~C.}\
  \bibnamefont {Halimeh}},\ }\href {\doibase 10.1103/PhysRevB.96.104436}
  {\bibfield  {journal} {\bibinfo  {journal} {Phys. Rev. B}\ }\textbf {\bibinfo
  {volume} {96}},\ \bibinfo {pages} {104436} (\bibinfo {year}
  {2017})}\BibitemShut {NoStop}%
\bibitem [{\citenamefont {Rossi}\ and\ \citenamefont
  {Dolcini}(2022)}]{rossi2022non}%
  \BibitemOpen
  \bibfield  {author} {\bibinfo {author} {\bibfnamefont {L.}~\bibnamefont
  {Rossi}}\ and\ \bibinfo {author} {\bibfnamefont {F.}~\bibnamefont
  {Dolcini}},\ }\href@noop {} {\bibfield  {journal} {\bibinfo  {journal} {arXiv
  preprint arXiv:2203.13874}\ } (\bibinfo {year} {2022})}\BibitemShut {NoStop}%
\bibitem [{\citenamefont {Mishra}\ \emph {et~al.}(2020)\citenamefont {Mishra},
  \citenamefont {Jafari},\ and\ \citenamefont {Akbari}}]{mishra2020disordered}%
  \BibitemOpen
  \bibfield  {author} {\bibinfo {author} {\bibfnamefont {U.}~\bibnamefont
  {Mishra}}, \bibinfo {author} {\bibfnamefont {R.}~\bibnamefont {Jafari}}, \
  and\ \bibinfo {author} {\bibfnamefont {A.}~\bibnamefont {Akbari}},\
  }\href@noop {} {\bibfield  {journal} {\bibinfo  {journal} {Journal of Physics
  A: Mathematical and Theoretical}\ }\textbf {\bibinfo {volume} {53}},\
  \bibinfo {pages} {375301} (\bibinfo {year} {2020})}\BibitemShut {NoStop}%
\bibitem [{\citenamefont {Sharma}\ \emph {et~al.}(2016)\citenamefont {Sharma},
  \citenamefont {Divakaran}, \citenamefont {Polkovnikov},\ and\ \citenamefont
  {Dutta}}]{SS}%
  \BibitemOpen
  \bibfield  {author} {\bibinfo {author} {\bibfnamefont {S.}~\bibnamefont
  {Sharma}}, \bibinfo {author} {\bibfnamefont {U.}~\bibnamefont {Divakaran}},
  \bibinfo {author} {\bibfnamefont {A.}~\bibnamefont {Polkovnikov}}, \ and\
  \bibinfo {author} {\bibfnamefont {A.}~\bibnamefont {Dutta}},\ }\href
  {\doibase 10.1103/PhysRevB.93.144306} {\bibfield  {journal} {\bibinfo
  {journal} {Phys. Rev. B}\ }\textbf {\bibinfo {volume} {93}},\ \bibinfo
  {pages} {144306} (\bibinfo {year} {2016})}\BibitemShut {NoStop}%
\bibitem [{\citenamefont {Sharma}\ \emph {et~al.}(2015)\citenamefont {Sharma},
  \citenamefont {Suzuki},\ and\ \citenamefont {Dutta}}]{PhysRevB.92.104306}%
  \BibitemOpen
  \bibfield  {author} {\bibinfo {author} {\bibfnamefont {S.}~\bibnamefont
  {Sharma}}, \bibinfo {author} {\bibfnamefont {S.}~\bibnamefont {Suzuki}}, \
  and\ \bibinfo {author} {\bibfnamefont {A.}~\bibnamefont {Dutta}},\ }\href
  {\doibase 10.1103/PhysRevB.92.104306} {\bibfield  {journal} {\bibinfo
  {journal} {Phys. Rev. B}\ }\textbf {\bibinfo {volume} {92}},\ \bibinfo
  {pages} {104306} (\bibinfo {year} {2015})}\BibitemShut {NoStop}%
\bibitem [{\citenamefont {Divakaran}\ \emph {et~al.}(2016)\citenamefont
  {Divakaran}, \citenamefont {Sharma},\ and\ \citenamefont
  {Dutta}}]{Divakaran16}%
  \BibitemOpen
  \bibfield  {author} {\bibinfo {author} {\bibfnamefont {U.}~\bibnamefont
  {Divakaran}}, \bibinfo {author} {\bibfnamefont {S.}~\bibnamefont {Sharma}}, \
  and\ \bibinfo {author} {\bibfnamefont {A.}~\bibnamefont {Dutta}},\ }\href
  {\doibase 10.1103/PhysRevE.93.052133} {\bibfield  {journal} {\bibinfo
  {journal} {Phys. Rev. E}\ }\textbf {\bibinfo {volume} {93}},\ \bibinfo
  {pages} {052133} (\bibinfo {year} {2016})}\BibitemShut {NoStop}%
\bibitem [{\citenamefont {Dutta}\ and\ \citenamefont {Dutta}(2017)}]{Dutta17}%
  \BibitemOpen
  \bibfield  {author} {\bibinfo {author} {\bibfnamefont {A.}~\bibnamefont
  {Dutta}}\ and\ \bibinfo {author} {\bibfnamefont {A.}~\bibnamefont {Dutta}},\
  }\href {\doibase 10.1103/PhysRevB.96.125113} {\bibfield  {journal} {\bibinfo
  {journal} {Phys. Rev. B}\ }\textbf {\bibinfo {volume} {96}},\ \bibinfo
  {pages} {125113} (\bibinfo {year} {2017})}\BibitemShut {NoStop}%
\bibitem [{\citenamefont {Vajna}\ and\ \citenamefont
  {D\'ora}(2015)}]{PhysRevB.91.155127}%
  \BibitemOpen
  \bibfield  {author} {\bibinfo {author} {\bibfnamefont {S.}~\bibnamefont
  {Vajna}}\ and\ \bibinfo {author} {\bibfnamefont {B.}~\bibnamefont {D\'ora}},\
  }\href {\doibase 10.1103/PhysRevB.91.155127} {\bibfield  {journal} {\bibinfo
  {journal} {Phys. Rev. B}\ }\textbf {\bibinfo {volume} {91}},\ \bibinfo
  {pages} {155127} (\bibinfo {year} {2015})}\BibitemShut {NoStop}%
\bibitem [{\citenamefont {Budich}\ and\ \citenamefont {Heyl}(2016)}]{Budich1}%
  \BibitemOpen
  \bibfield  {author} {\bibinfo {author} {\bibfnamefont {J.~C.}\ \bibnamefont
  {Budich}}\ and\ \bibinfo {author} {\bibfnamefont {M.}~\bibnamefont {Heyl}},\
  }\href {\doibase 10.1103/PhysRevB.93.085416} {\bibfield  {journal} {\bibinfo
  {journal} {Phys. Rev. B}\ }\textbf {\bibinfo {volume} {93}},\ \bibinfo
  {pages} {085416} (\bibinfo {year} {2016})}\BibitemShut {NoStop}%
\bibitem [{\citenamefont {Vajna}\ and\ \citenamefont
  {D\'ora}(2014{\natexlab{b}})}]{PhysRevB.89.161105}%
  \BibitemOpen
  \bibfield  {author} {\bibinfo {author} {\bibfnamefont {S.}~\bibnamefont
  {Vajna}}\ and\ \bibinfo {author} {\bibfnamefont {B.}~\bibnamefont {D\'ora}},\
  }\href {\doibase 10.1103/PhysRevB.89.161105} {\bibfield  {journal} {\bibinfo
  {journal} {Phys. Rev. B}\ }\textbf {\bibinfo {volume} {89}},\ \bibinfo
  {pages} {161105} (\bibinfo {year} {2014}{\natexlab{b}})}\BibitemShut
  {NoStop}%
\bibitem [{\citenamefont {Palmai}(2015)}]{PhysRevB.92.235433}%
  \BibitemOpen
  \bibfield  {author} {\bibinfo {author} {\bibfnamefont {T.}~\bibnamefont
  {Palmai}},\ }\href {\doibase 10.1103/PhysRevB.92.235433} {\bibfield
  {journal} {\bibinfo  {journal} {Phys. Rev. B}\ }\textbf {\bibinfo {volume}
  {92}},\ \bibinfo {pages} {235433} (\bibinfo {year} {2015})}\BibitemShut
  {NoStop}%
\bibitem [{\citenamefont {Andraschko}\ and\ \citenamefont
  {Sirker}(2014)}]{PhysRevB.89.125120}%
  \BibitemOpen
  \bibfield  {author} {\bibinfo {author} {\bibfnamefont {F.}~\bibnamefont
  {Andraschko}}\ and\ \bibinfo {author} {\bibfnamefont {J.}~\bibnamefont
  {Sirker}},\ }\href {\doibase 10.1103/PhysRevB.89.125120} {\bibfield
  {journal} {\bibinfo  {journal} {Phys. Rev. B}\ }\textbf {\bibinfo {volume}
  {89}},\ \bibinfo {pages} {125120} (\bibinfo {year} {2014})}\BibitemShut
  {NoStop}%
\bibitem [{\citenamefont {Modak}\ and\ \citenamefont
  {Rakshit}(2021)}]{Modak21}%
  \BibitemOpen
  \bibfield  {author} {\bibinfo {author} {\bibfnamefont {R.}~\bibnamefont
  {Modak}}\ and\ \bibinfo {author} {\bibfnamefont {D.}~\bibnamefont
  {Rakshit}},\ }\href {\doibase 10.1103/PhysRevB.103.224310} {\bibfield
  {journal} {\bibinfo  {journal} {Phys. Rev. B}\ }\textbf {\bibinfo {volume}
  {103}},\ \bibinfo {pages} {224310} (\bibinfo {year} {2021})}\BibitemShut
  {NoStop}%
\bibitem [{\citenamefont {Abdi}(2019)}]{Abdi19}%
  \BibitemOpen
  \bibfield  {author} {\bibinfo {author} {\bibfnamefont {M.}~\bibnamefont
  {Abdi}},\ }\href {\doibase 10.1103/PhysRevB.100.184310} {\bibfield  {journal}
  {\bibinfo  {journal} {Phys. Rev. B}\ }\textbf {\bibinfo {volume} {100}},\
  \bibinfo {pages} {184310} (\bibinfo {year} {2019})}\BibitemShut {NoStop}%
\bibitem [{\citenamefont {Syed}\ \emph {et~al.}(2021)\citenamefont {Syed},
  \citenamefont {Enss},\ and\ \citenamefont {Defenu}}]{PhysRevB.103.064306}%
  \BibitemOpen
  \bibfield  {author} {\bibinfo {author} {\bibfnamefont {M.}~\bibnamefont
  {Syed}}, \bibinfo {author} {\bibfnamefont {T.}~\bibnamefont {Enss}}, \ and\
  \bibinfo {author} {\bibfnamefont {N.}~\bibnamefont {Defenu}},\ }\href
  {\doibase 10.1103/PhysRevB.103.064306} {\bibfield  {journal} {\bibinfo
  {journal} {Phys. Rev. B}\ }\textbf {\bibinfo {volume} {103}},\ \bibinfo
  {pages} {064306} (\bibinfo {year} {2021})}\BibitemShut {NoStop}%
\bibitem [{\citenamefont {Stumper}\ \emph {et~al.}(2022)\citenamefont
  {Stumper}, \citenamefont {Thoss},\ and\ \citenamefont
  {Okamoto}}]{PhysRevResearch.4.013002}%
  \BibitemOpen
  \bibfield  {author} {\bibinfo {author} {\bibfnamefont {S.}~\bibnamefont
  {Stumper}}, \bibinfo {author} {\bibfnamefont {M.}~\bibnamefont {Thoss}}, \
  and\ \bibinfo {author} {\bibfnamefont {J.}~\bibnamefont {Okamoto}},\ }\href
  {\doibase 10.1103/PhysRevResearch.4.013002} {\bibfield  {journal} {\bibinfo
  {journal} {Phys. Rev. Research}\ }\textbf {\bibinfo {volume} {4}},\ \bibinfo
  {pages} {013002} (\bibinfo {year} {2022})}\BibitemShut {NoStop}%
\bibitem [{\citenamefont {Zamani}\ \emph {et~al.}(2020)\citenamefont {Zamani},
  \citenamefont {Jafari},\ and\ \citenamefont {Langari}}]{Zamani20}%
  \BibitemOpen
  \bibfield  {author} {\bibinfo {author} {\bibfnamefont {S.}~\bibnamefont
  {Zamani}}, \bibinfo {author} {\bibfnamefont {R.}~\bibnamefont {Jafari}}, \
  and\ \bibinfo {author} {\bibfnamefont {A.}~\bibnamefont {Langari}},\ }\href
  {\doibase 10.1103/PhysRevB.102.144306} {\bibfield  {journal} {\bibinfo
  {journal} {Phys. Rev. B}\ }\textbf {\bibinfo {volume} {102}},\ \bibinfo
  {pages} {144306} (\bibinfo {year} {2020})}\BibitemShut {NoStop}%
\bibitem [{\citenamefont {Jafari}\ \emph {et~al.}(2022)\citenamefont {Jafari},
  \citenamefont {Akbari}, \citenamefont {Mishra},\ and\ \citenamefont
  {Johannesson}}]{Jafari22}%
  \BibitemOpen
  \bibfield  {author} {\bibinfo {author} {\bibfnamefont {R.}~\bibnamefont
  {Jafari}}, \bibinfo {author} {\bibfnamefont {A.}~\bibnamefont {Akbari}},
  \bibinfo {author} {\bibfnamefont {U.}~\bibnamefont {Mishra}}, \ and\ \bibinfo
  {author} {\bibfnamefont {H.}~\bibnamefont {Johannesson}},\ }\href {\doibase
  10.1103/PhysRevB.105.094311} {\bibfield  {journal} {\bibinfo  {journal}
  {Phys. Rev. B}\ }\textbf {\bibinfo {volume} {105}},\ \bibinfo {pages}
  {094311} (\bibinfo {year} {2022})}\BibitemShut {NoStop}%
\bibitem [{\citenamefont {Jafari}\ and\ \citenamefont
  {Akbari}(2021{\natexlab{a}})}]{Jafari21a}%
  \BibitemOpen
  \bibfield  {author} {\bibinfo {author} {\bibfnamefont {R.}~\bibnamefont
  {Jafari}}\ and\ \bibinfo {author} {\bibfnamefont {A.}~\bibnamefont
  {Akbari}},\ }\href {\doibase 10.1103/PhysRevA.103.012204} {\bibfield
  {journal} {\bibinfo  {journal} {Phys. Rev. A}\ }\textbf {\bibinfo {volume}
  {103}},\ \bibinfo {pages} {012204} (\bibinfo {year}
  {2021}{\natexlab{a}})}\BibitemShut {NoStop}%
\bibitem [{\citenamefont {Jurcevic}\ \emph {et~al.}(2017)\citenamefont
  {Jurcevic}, \citenamefont {Shen}, \citenamefont {Hauke}, \citenamefont
  {Maier}, \citenamefont {Brydges}, \citenamefont {Hempel}, \citenamefont
  {Lanyon}, \citenamefont {Heyl}, \citenamefont {Blatt},\ and\ \citenamefont
  {Roos}}]{PhysRevLett.119.080501}%
  \BibitemOpen
  \bibfield  {author} {\bibinfo {author} {\bibfnamefont {P.}~\bibnamefont
  {Jurcevic}}, \bibinfo {author} {\bibfnamefont {H.}~\bibnamefont {Shen}},
  \bibinfo {author} {\bibfnamefont {P.}~\bibnamefont {Hauke}}, \bibinfo
  {author} {\bibfnamefont {C.}~\bibnamefont {Maier}}, \bibinfo {author}
  {\bibfnamefont {T.}~\bibnamefont {Brydges}}, \bibinfo {author} {\bibfnamefont
  {C.}~\bibnamefont {Hempel}}, \bibinfo {author} {\bibfnamefont {B.~P.}\
  \bibnamefont {Lanyon}}, \bibinfo {author} {\bibfnamefont {M.}~\bibnamefont
  {Heyl}}, \bibinfo {author} {\bibfnamefont {R.}~\bibnamefont {Blatt}}, \ and\
  \bibinfo {author} {\bibfnamefont {C.~F.}\ \bibnamefont {Roos}},\ }\href
  {\doibase 10.1103/PhysRevLett.119.080501} {\bibfield  {journal} {\bibinfo
  {journal} {Phys. Rev. Lett.}\ }\textbf {\bibinfo {volume} {119}},\ \bibinfo
  {pages} {080501} (\bibinfo {year} {2017})}\BibitemShut {NoStop}%
\bibitem [{\citenamefont {Nie}\ \emph {et~al.}(2020)\citenamefont {Nie},
  \citenamefont {Wei}, \citenamefont {Chen}, \citenamefont {Zhang},
  \citenamefont {Zhao}, \citenamefont {Qiu}, \citenamefont {Tian},
  \citenamefont {Ji}, \citenamefont {Xin}, \citenamefont {Lu},\ and\
  \citenamefont {Li}}]{Nie20}%
  \BibitemOpen
  \bibfield  {author} {\bibinfo {author} {\bibfnamefont {X.}~\bibnamefont
  {Nie}}, \bibinfo {author} {\bibfnamefont {B.-B.}\ \bibnamefont {Wei}},
  \bibinfo {author} {\bibfnamefont {X.}~\bibnamefont {Chen}}, \bibinfo {author}
  {\bibfnamefont {Z.}~\bibnamefont {Zhang}}, \bibinfo {author} {\bibfnamefont
  {X.}~\bibnamefont {Zhao}}, \bibinfo {author} {\bibfnamefont {C.}~\bibnamefont
  {Qiu}}, \bibinfo {author} {\bibfnamefont {Y.}~\bibnamefont {Tian}}, \bibinfo
  {author} {\bibfnamefont {Y.}~\bibnamefont {Ji}}, \bibinfo {author}
  {\bibfnamefont {T.}~\bibnamefont {Xin}}, \bibinfo {author} {\bibfnamefont
  {D.}~\bibnamefont {Lu}}, \ and\ \bibinfo {author} {\bibfnamefont
  {J.}~\bibnamefont {Li}},\ }\href {\doibase 10.1103/PhysRevLett.124.250601}
  {\bibfield  {journal} {\bibinfo  {journal} {Phys. Rev. Lett.}\ }\textbf
  {\bibinfo {volume} {124}},\ \bibinfo {pages} {250601} (\bibinfo {year}
  {2020})}\BibitemShut {NoStop}%
\bibitem [{\citenamefont {Fl{\"a}schner}\ \emph {et~al.}(2018)\citenamefont
  {Fl{\"a}schner}, \citenamefont {Vogel}, \citenamefont {Tarnowski},
  \citenamefont {Rem}, \citenamefont {L{\"u}hmann}, \citenamefont {Heyl},
  \citenamefont {Budich}, \citenamefont {Mathey}, \citenamefont {Sengstock},\
  and\ \citenamefont {Weitenberg}}]{flaschner2018observation}%
  \BibitemOpen
  \bibfield  {author} {\bibinfo {author} {\bibfnamefont {N.}~\bibnamefont
  {Fl{\"a}schner}}, \bibinfo {author} {\bibfnamefont {D.}~\bibnamefont
  {Vogel}}, \bibinfo {author} {\bibfnamefont {M.}~\bibnamefont {Tarnowski}},
  \bibinfo {author} {\bibfnamefont {B.}~\bibnamefont {Rem}}, \bibinfo {author}
  {\bibfnamefont {D.-S.}\ \bibnamefont {L{\"u}hmann}}, \bibinfo {author}
  {\bibfnamefont {M.}~\bibnamefont {Heyl}}, \bibinfo {author} {\bibfnamefont
  {J.}~\bibnamefont {Budich}}, \bibinfo {author} {\bibfnamefont
  {L.}~\bibnamefont {Mathey}}, \bibinfo {author} {\bibfnamefont
  {K.}~\bibnamefont {Sengstock}}, \ and\ \bibinfo {author} {\bibfnamefont
  {C.}~\bibnamefont {Weitenberg}},\ }\href@noop {} {\bibfield  {journal}
  {\bibinfo  {journal} {Nature Physics}\ }\textbf {\bibinfo {volume} {14}},\
  \bibinfo {pages} {265} (\bibinfo {year} {2018})}\BibitemShut {NoStop}%
\bibitem [{\citenamefont {Nag}\ \emph {et~al.}(2012)\citenamefont {Nag},
  \citenamefont {Divakaran},\ and\ \citenamefont {Dutta}}]{Nag12}%
  \BibitemOpen
  \bibfield  {author} {\bibinfo {author} {\bibfnamefont {T.}~\bibnamefont
  {Nag}}, \bibinfo {author} {\bibfnamefont {U.}~\bibnamefont {Divakaran}}, \
  and\ \bibinfo {author} {\bibfnamefont {A.}~\bibnamefont {Dutta}},\ }\href
  {\doibase 10.1103/PhysRevB.86.020401} {\bibfield  {journal} {\bibinfo
  {journal} {Phys. Rev. B}\ }\textbf {\bibinfo {volume} {86}},\ \bibinfo
  {pages} {020401} (\bibinfo {year} {2012})}\BibitemShut {NoStop}%
\bibitem [{\citenamefont {Sachdeva}\ \emph {et~al.}(2014)\citenamefont
  {Sachdeva}, \citenamefont {Nag}, \citenamefont {Agarwal},\ and\ \citenamefont
  {Dutta}}]{Sachdeva14}%
  \BibitemOpen
  \bibfield  {author} {\bibinfo {author} {\bibfnamefont {R.}~\bibnamefont
  {Sachdeva}}, \bibinfo {author} {\bibfnamefont {T.}~\bibnamefont {Nag}},
  \bibinfo {author} {\bibfnamefont {A.}~\bibnamefont {Agarwal}}, \ and\
  \bibinfo {author} {\bibfnamefont {A.}~\bibnamefont {Dutta}},\ }\href
  {\doibase 10.1103/PhysRevB.90.045421} {\bibfield  {journal} {\bibinfo
  {journal} {Phys. Rev. B}\ }\textbf {\bibinfo {volume} {90}},\ \bibinfo
  {pages} {045421} (\bibinfo {year} {2014})}\BibitemShut {NoStop}%
\bibitem [{\citenamefont {Nag}(2016)}]{Nag16}%
  \BibitemOpen
  \bibfield  {author} {\bibinfo {author} {\bibfnamefont {T.}~\bibnamefont
  {Nag}},\ }\href {\doibase 10.1103/PhysRevE.93.062119} {\bibfield  {journal}
  {\bibinfo  {journal} {Phys. Rev. E}\ }\textbf {\bibinfo {volume} {93}},\
  \bibinfo {pages} {062119} (\bibinfo {year} {2016})}\BibitemShut {NoStop}%
\bibitem [{\citenamefont {Suzuki}\ \emph {et~al.}(2016)\citenamefont {Suzuki},
  \citenamefont {Nag},\ and\ \citenamefont {Dutta}}]{Suzuki16}%
  \BibitemOpen
  \bibfield  {author} {\bibinfo {author} {\bibfnamefont {S.}~\bibnamefont
  {Suzuki}}, \bibinfo {author} {\bibfnamefont {T.}~\bibnamefont {Nag}}, \ and\
  \bibinfo {author} {\bibfnamefont {A.}~\bibnamefont {Dutta}},\ }\href
  {\doibase 10.1103/PhysRevA.93.012112} {\bibfield  {journal} {\bibinfo
  {journal} {Phys. Rev. A}\ }\textbf {\bibinfo {volume} {93}},\ \bibinfo
  {pages} {012112} (\bibinfo {year} {2016})}\BibitemShut {NoStop}%
\bibitem [{\citenamefont {Quan}\ \emph {et~al.}(2006)\citenamefont {Quan},
  \citenamefont {Song}, \citenamefont {Liu}, \citenamefont {Zanardi},\ and\
  \citenamefont {Sun}}]{PhysRevLett.96.140604}%
  \BibitemOpen
  \bibfield  {author} {\bibinfo {author} {\bibfnamefont {H.~T.}\ \bibnamefont
  {Quan}}, \bibinfo {author} {\bibfnamefont {Z.}~\bibnamefont {Song}}, \bibinfo
  {author} {\bibfnamefont {X.~F.}\ \bibnamefont {Liu}}, \bibinfo {author}
  {\bibfnamefont {P.}~\bibnamefont {Zanardi}}, \ and\ \bibinfo {author}
  {\bibfnamefont {C.~P.}\ \bibnamefont {Sun}},\ }\href {\doibase
  10.1103/PhysRevLett.96.140604} {\bibfield  {journal} {\bibinfo  {journal}
  {Phys. Rev. Lett.}\ }\textbf {\bibinfo {volume} {96}},\ \bibinfo {pages}
  {140604} (\bibinfo {year} {2006})}\BibitemShut {NoStop}%
\bibitem [{\citenamefont {Cucchietti}\ \emph {et~al.}(2003)\citenamefont
  {Cucchietti}, \citenamefont {Dalvit}, \citenamefont {Paz},\ and\
  \citenamefont {Zurek}}]{Cucchietti03}%
  \BibitemOpen
  \bibfield  {author} {\bibinfo {author} {\bibfnamefont {F.~M.}\ \bibnamefont
  {Cucchietti}}, \bibinfo {author} {\bibfnamefont {D.~A.~R.}\ \bibnamefont
  {Dalvit}}, \bibinfo {author} {\bibfnamefont {J.~P.}\ \bibnamefont {Paz}}, \
  and\ \bibinfo {author} {\bibfnamefont {W.~H.}\ \bibnamefont {Zurek}},\ }\href
  {\doibase 10.1103/PhysRevLett.91.210403} {\bibfield  {journal} {\bibinfo
  {journal} {Phys. Rev. Lett.}\ }\textbf {\bibinfo {volume} {91}},\ \bibinfo
  {pages} {210403} (\bibinfo {year} {2003})}\BibitemShut {NoStop}%
\bibitem [{\citenamefont {Jafari}\ and\ \citenamefont
  {Johannesson}(2017{\natexlab{a}})}]{Jafari17b}%
  \BibitemOpen
  \bibfield  {author} {\bibinfo {author} {\bibfnamefont {R.}~\bibnamefont
  {Jafari}}\ and\ \bibinfo {author} {\bibfnamefont {H.}~\bibnamefont
  {Johannesson}},\ }\href {\doibase 10.1103/PhysRevB.96.224302} {\bibfield
  {journal} {\bibinfo  {journal} {Phys. Rev. B}\ }\textbf {\bibinfo {volume}
  {96}},\ \bibinfo {pages} {224302} (\bibinfo {year}
  {2017}{\natexlab{a}})}\BibitemShut {NoStop}%
\bibitem [{\citenamefont {Kosior}\ \emph {et~al.}(2018)\citenamefont {Kosior},
  \citenamefont {Syrwid},\ and\ \citenamefont {Sacha}}]{Kosior18}%
  \BibitemOpen
  \bibfield  {author} {\bibinfo {author} {\bibfnamefont {A.}~\bibnamefont
  {Kosior}}, \bibinfo {author} {\bibfnamefont {A.}~\bibnamefont {Syrwid}}, \
  and\ \bibinfo {author} {\bibfnamefont {K.}~\bibnamefont {Sacha}},\ }\href
  {\doibase 10.1103/PhysRevA.98.023612} {\bibfield  {journal} {\bibinfo
  {journal} {Phys. Rev. A}\ }\textbf {\bibinfo {volume} {98}},\ \bibinfo
  {pages} {023612} (\bibinfo {year} {2018})}\BibitemShut {NoStop}%
\bibitem [{\citenamefont {Kosior}\ and\ \citenamefont
  {Sacha}(2018)}]{Kosior18b}%
  \BibitemOpen
  \bibfield  {author} {\bibinfo {author} {\bibfnamefont {A.}~\bibnamefont
  {Kosior}}\ and\ \bibinfo {author} {\bibfnamefont {K.}~\bibnamefont {Sacha}},\
  }\href {\doibase 10.1103/PhysRevA.97.053621} {\bibfield  {journal} {\bibinfo
  {journal} {Phys. Rev. A}\ }\textbf {\bibinfo {volume} {97}},\ \bibinfo
  {pages} {053621} (\bibinfo {year} {2018})}\BibitemShut {NoStop}%
\bibitem [{\citenamefont {Jafari}\ and\ \citenamefont
  {Akbari}(2021{\natexlab{b}})}]{Jafari21}%
  \BibitemOpen
  \bibfield  {author} {\bibinfo {author} {\bibfnamefont {R.}~\bibnamefont
  {Jafari}}\ and\ \bibinfo {author} {\bibfnamefont {A.}~\bibnamefont
  {Akbari}},\ }\href {\doibase 10.1103/PhysRevA.103.012204} {\bibfield
  {journal} {\bibinfo  {journal} {Phys. Rev. A}\ }\textbf {\bibinfo {volume}
  {103}},\ \bibinfo {pages} {012204} (\bibinfo {year}
  {2021}{\natexlab{b}})}\BibitemShut {NoStop}%
\bibitem [{\citenamefont {Jafari}\ and\ \citenamefont
  {Johannesson}(2017{\natexlab{b}})}]{Jafari17}%
  \BibitemOpen
  \bibfield  {author} {\bibinfo {author} {\bibfnamefont {R.}~\bibnamefont
  {Jafari}}\ and\ \bibinfo {author} {\bibfnamefont {H.}~\bibnamefont
  {Johannesson}},\ }\href {\doibase 10.1103/PhysRevLett.118.015701} {\bibfield
  {journal} {\bibinfo  {journal} {Phys. Rev. Lett.}\ }\textbf {\bibinfo
  {volume} {118}},\ \bibinfo {pages} {015701} (\bibinfo {year}
  {2017}{\natexlab{b}})}\BibitemShut {NoStop}%
\bibitem [{\citenamefont {Zhou}\ and\ \citenamefont
  {Du}(2021{\natexlab{a}})}]{zhou2021floquet}%
  \BibitemOpen
  \bibfield  {author} {\bibinfo {author} {\bibfnamefont {L.}~\bibnamefont
  {Zhou}}\ and\ \bibinfo {author} {\bibfnamefont {Q.}~\bibnamefont {Du}},\
  }\href@noop {} {\bibfield  {journal} {\bibinfo  {journal} {Journal of
  Physics: Condensed Matter}\ }\textbf {\bibinfo {volume} {33}},\ \bibinfo
  {pages} {345403} (\bibinfo {year} {2021}{\natexlab{a}})}\BibitemShut
  {NoStop}%
\bibitem [{\citenamefont {Yang}\ \emph {et~al.}(2019)\citenamefont {Yang},
  \citenamefont {Zhou}, \citenamefont {Ma}, \citenamefont {Kong}, \citenamefont
  {Wang}, \citenamefont {Qin}, \citenamefont {Rong}, \citenamefont {Wang},
  \citenamefont {Shi}, \citenamefont {Gong},\ and\ \citenamefont
  {Du}}]{Yang19}%
  \BibitemOpen
  \bibfield  {author} {\bibinfo {author} {\bibfnamefont {K.}~\bibnamefont
  {Yang}}, \bibinfo {author} {\bibfnamefont {L.}~\bibnamefont {Zhou}}, \bibinfo
  {author} {\bibfnamefont {W.}~\bibnamefont {Ma}}, \bibinfo {author}
  {\bibfnamefont {X.}~\bibnamefont {Kong}}, \bibinfo {author} {\bibfnamefont
  {P.}~\bibnamefont {Wang}}, \bibinfo {author} {\bibfnamefont {X.}~\bibnamefont
  {Qin}}, \bibinfo {author} {\bibfnamefont {X.}~\bibnamefont {Rong}}, \bibinfo
  {author} {\bibfnamefont {Y.}~\bibnamefont {Wang}}, \bibinfo {author}
  {\bibfnamefont {F.}~\bibnamefont {Shi}}, \bibinfo {author} {\bibfnamefont
  {J.}~\bibnamefont {Gong}}, \ and\ \bibinfo {author} {\bibfnamefont
  {J.}~\bibnamefont {Du}},\ }\href {\doibase 10.1103/PhysRevB.100.085308}
  {\bibfield  {journal} {\bibinfo  {journal} {Phys. Rev. B}\ }\textbf {\bibinfo
  {volume} {100}},\ \bibinfo {pages} {085308} (\bibinfo {year}
  {2019})}\BibitemShut {NoStop}%
\bibitem [{\citenamefont {Bergholtz}\ and\ \citenamefont
  {Budich}(2019)}]{Bergholtz19}%
  \BibitemOpen
  \bibfield  {author} {\bibinfo {author} {\bibfnamefont {E.~J.}\ \bibnamefont
  {Bergholtz}}\ and\ \bibinfo {author} {\bibfnamefont {J.~C.}\ \bibnamefont
  {Budich}},\ }\href {\doibase 10.1103/PhysRevResearch.1.012003} {\bibfield
  {journal} {\bibinfo  {journal} {Phys. Rev. Research}\ }\textbf {\bibinfo
  {volume} {1}},\ \bibinfo {pages} {012003} (\bibinfo {year}
  {2019})}\BibitemShut {NoStop}%
\bibitem [{\citenamefont {Yang}\ \emph {et~al.}(2021)\citenamefont {Yang},
  \citenamefont {Morampudi},\ and\ \citenamefont {Bergholtz}}]{Yang21}%
  \BibitemOpen
  \bibfield  {author} {\bibinfo {author} {\bibfnamefont {K.}~\bibnamefont
  {Yang}}, \bibinfo {author} {\bibfnamefont {S.~C.}\ \bibnamefont {Morampudi}},
  \ and\ \bibinfo {author} {\bibfnamefont {E.~J.}\ \bibnamefont {Bergholtz}},\
  }\href {\doibase 10.1103/PhysRevLett.126.077201} {\bibfield  {journal}
  {\bibinfo  {journal} {Phys. Rev. Lett.}\ }\textbf {\bibinfo {volume} {126}},\
  \bibinfo {pages} {077201} (\bibinfo {year} {2021})}\BibitemShut {NoStop}%
\bibitem [{\citenamefont {Kozii}\ and\ \citenamefont
  {Fu}(2017)}]{kozii2017non}%
  \BibitemOpen
  \bibfield  {author} {\bibinfo {author} {\bibfnamefont {V.}~\bibnamefont
  {Kozii}}\ and\ \bibinfo {author} {\bibfnamefont {L.}~\bibnamefont {Fu}},\
  }\href@noop {} {\bibfield  {journal} {\bibinfo  {journal} {arXiv preprint
  arXiv:1708.05841}\ } (\bibinfo {year} {2017})}\BibitemShut {NoStop}%
\bibitem [{\citenamefont {Yoshida}\ \emph {et~al.}(2018)\citenamefont
  {Yoshida}, \citenamefont {Peters},\ and\ \citenamefont
  {Kawakami}}]{Yoshida18}%
  \BibitemOpen
  \bibfield  {author} {\bibinfo {author} {\bibfnamefont {T.}~\bibnamefont
  {Yoshida}}, \bibinfo {author} {\bibfnamefont {R.}~\bibnamefont {Peters}}, \
  and\ \bibinfo {author} {\bibfnamefont {N.}~\bibnamefont {Kawakami}},\ }\href
  {\doibase 10.1103/PhysRevB.98.035141} {\bibfield  {journal} {\bibinfo
  {journal} {Phys. Rev. B}\ }\textbf {\bibinfo {volume} {98}},\ \bibinfo
  {pages} {035141} (\bibinfo {year} {2018})}\BibitemShut {NoStop}%
\bibitem [{\citenamefont {Shen}\ \emph {et~al.}(2018)\citenamefont {Shen},
  \citenamefont {Zhen},\ and\ \citenamefont {Fu}}]{Shen18}%
  \BibitemOpen
  \bibfield  {author} {\bibinfo {author} {\bibfnamefont {H.}~\bibnamefont
  {Shen}}, \bibinfo {author} {\bibfnamefont {B.}~\bibnamefont {Zhen}}, \ and\
  \bibinfo {author} {\bibfnamefont {L.}~\bibnamefont {Fu}},\ }\href {\doibase
  10.1103/PhysRevLett.120.146402} {\bibfield  {journal} {\bibinfo  {journal}
  {Phys. Rev. Lett.}\ }\textbf {\bibinfo {volume} {120}},\ \bibinfo {pages}
  {146402} (\bibinfo {year} {2018})}\BibitemShut {NoStop}%
\bibitem [{\citenamefont {Gou}\ \emph {et~al.}(2020)\citenamefont {Gou},
  \citenamefont {Chen}, \citenamefont {Xie}, \citenamefont {Xiao},
  \citenamefont {Deng}, \citenamefont {Gadway}, \citenamefont {Yi},\ and\
  \citenamefont {Yan}}]{Gou20}%
  \BibitemOpen
  \bibfield  {author} {\bibinfo {author} {\bibfnamefont {W.}~\bibnamefont
  {Gou}}, \bibinfo {author} {\bibfnamefont {T.}~\bibnamefont {Chen}}, \bibinfo
  {author} {\bibfnamefont {D.}~\bibnamefont {Xie}}, \bibinfo {author}
  {\bibfnamefont {T.}~\bibnamefont {Xiao}}, \bibinfo {author} {\bibfnamefont
  {T.-S.}\ \bibnamefont {Deng}}, \bibinfo {author} {\bibfnamefont
  {B.}~\bibnamefont {Gadway}}, \bibinfo {author} {\bibfnamefont
  {W.}~\bibnamefont {Yi}}, \ and\ \bibinfo {author} {\bibfnamefont
  {B.}~\bibnamefont {Yan}},\ }\href {\doibase 10.1103/PhysRevLett.124.070402}
  {\bibfield  {journal} {\bibinfo  {journal} {Phys. Rev. Lett.}\ }\textbf
  {\bibinfo {volume} {124}},\ \bibinfo {pages} {070402} (\bibinfo {year}
  {2020})}\BibitemShut {NoStop}%
\bibitem [{\citenamefont {Li}\ \emph {et~al.}(2019)\citenamefont {Li},
  \citenamefont {Harter}, \citenamefont {Liu}, \citenamefont {de~Melo},
  \citenamefont {Joglekar},\ and\ \citenamefont {Luo}}]{li2019observation}%
  \BibitemOpen
  \bibfield  {author} {\bibinfo {author} {\bibfnamefont {J.}~\bibnamefont
  {Li}}, \bibinfo {author} {\bibfnamefont {A.~K.}\ \bibnamefont {Harter}},
  \bibinfo {author} {\bibfnamefont {J.}~\bibnamefont {Liu}}, \bibinfo {author}
  {\bibfnamefont {L.}~\bibnamefont {de~Melo}}, \bibinfo {author} {\bibfnamefont
  {Y.~N.}\ \bibnamefont {Joglekar}}, \ and\ \bibinfo {author} {\bibfnamefont
  {L.}~\bibnamefont {Luo}},\ }\href@noop {} {\bibfield  {journal} {\bibinfo
  {journal} {Nature communications}\ }\textbf {\bibinfo {volume} {10}},\
  \bibinfo {pages} {855} (\bibinfo {year} {2019})}\BibitemShut {NoStop}%
\bibitem [{\citenamefont {Zeuner}\ \emph {et~al.}(2015)\citenamefont {Zeuner},
  \citenamefont {Rechtsman}, \citenamefont {Plotnik}, \citenamefont {Lumer},
  \citenamefont {Nolte}, \citenamefont {Rudner}, \citenamefont {Segev},\ and\
  \citenamefont {Szameit}}]{Zeuner15}%
  \BibitemOpen
  \bibfield  {author} {\bibinfo {author} {\bibfnamefont {J.~M.}\ \bibnamefont
  {Zeuner}}, \bibinfo {author} {\bibfnamefont {M.~C.}\ \bibnamefont
  {Rechtsman}}, \bibinfo {author} {\bibfnamefont {Y.}~\bibnamefont {Plotnik}},
  \bibinfo {author} {\bibfnamefont {Y.}~\bibnamefont {Lumer}}, \bibinfo
  {author} {\bibfnamefont {S.}~\bibnamefont {Nolte}}, \bibinfo {author}
  {\bibfnamefont {M.~S.}\ \bibnamefont {Rudner}}, \bibinfo {author}
  {\bibfnamefont {M.}~\bibnamefont {Segev}}, \ and\ \bibinfo {author}
  {\bibfnamefont {A.}~\bibnamefont {Szameit}},\ }\href {\doibase
  10.1103/PhysRevLett.115.040402} {\bibfield  {journal} {\bibinfo  {journal}
  {Phys. Rev. Lett.}\ }\textbf {\bibinfo {volume} {115}},\ \bibinfo {pages}
  {040402} (\bibinfo {year} {2015})}\BibitemShut {NoStop}%
\bibitem [{\citenamefont {Weimann}\ \emph {et~al.}(2017)\citenamefont
  {Weimann}, \citenamefont {Kremer}, \citenamefont {Plotnik}, \citenamefont
  {Lumer}, \citenamefont {Nolte}, \citenamefont {Makris}, \citenamefont
  {Segev}, \citenamefont {Rechtsman},\ and\ \citenamefont
  {Szameit}}]{weimann2017topologically}%
  \BibitemOpen
  \bibfield  {author} {\bibinfo {author} {\bibfnamefont {S.}~\bibnamefont
  {Weimann}}, \bibinfo {author} {\bibfnamefont {M.}~\bibnamefont {Kremer}},
  \bibinfo {author} {\bibfnamefont {Y.}~\bibnamefont {Plotnik}}, \bibinfo
  {author} {\bibfnamefont {Y.}~\bibnamefont {Lumer}}, \bibinfo {author}
  {\bibfnamefont {S.}~\bibnamefont {Nolte}}, \bibinfo {author} {\bibfnamefont
  {K.~G.}\ \bibnamefont {Makris}}, \bibinfo {author} {\bibfnamefont
  {M.}~\bibnamefont {Segev}}, \bibinfo {author} {\bibfnamefont {M.~C.}\
  \bibnamefont {Rechtsman}}, \ and\ \bibinfo {author} {\bibfnamefont
  {A.}~\bibnamefont {Szameit}},\ }\href@noop {} {\bibfield  {journal} {\bibinfo
   {journal} {Nature materials}\ }\textbf {\bibinfo {volume} {16}},\ \bibinfo
  {pages} {433} (\bibinfo {year} {2017})}\BibitemShut {NoStop}%
\bibitem [{\citenamefont {Zhu}\ \emph {et~al.}(2018)\citenamefont {Zhu},
  \citenamefont {Fang}, \citenamefont {Li}, \citenamefont {Sun}, \citenamefont
  {Li}, \citenamefont {Jing},\ and\ \citenamefont {Chen}}]{Weiwei18}%
  \BibitemOpen
  \bibfield  {author} {\bibinfo {author} {\bibfnamefont {W.}~\bibnamefont
  {Zhu}}, \bibinfo {author} {\bibfnamefont {X.}~\bibnamefont {Fang}}, \bibinfo
  {author} {\bibfnamefont {D.}~\bibnamefont {Li}}, \bibinfo {author}
  {\bibfnamefont {Y.}~\bibnamefont {Sun}}, \bibinfo {author} {\bibfnamefont
  {Y.}~\bibnamefont {Li}}, \bibinfo {author} {\bibfnamefont {Y.}~\bibnamefont
  {Jing}}, \ and\ \bibinfo {author} {\bibfnamefont {H.}~\bibnamefont {Chen}},\
  }\href {\doibase 10.1103/PhysRevLett.121.124501} {\bibfield  {journal}
  {\bibinfo  {journal} {Phys. Rev. Lett.}\ }\textbf {\bibinfo {volume} {121}},\
  \bibinfo {pages} {124501} (\bibinfo {year} {2018})}\BibitemShut {NoStop}%
\bibitem [{\citenamefont {Gao}\ \emph {et~al.}(2020)\citenamefont {Gao},
  \citenamefont {Xue}, \citenamefont {Wang}, \citenamefont {Gu}, \citenamefont
  {Liu}, \citenamefont {Zhu},\ and\ \citenamefont {Zhang}}]{Gao20}%
  \BibitemOpen
  \bibfield  {author} {\bibinfo {author} {\bibfnamefont {H.}~\bibnamefont
  {Gao}}, \bibinfo {author} {\bibfnamefont {H.}~\bibnamefont {Xue}}, \bibinfo
  {author} {\bibfnamefont {Q.}~\bibnamefont {Wang}}, \bibinfo {author}
  {\bibfnamefont {Z.}~\bibnamefont {Gu}}, \bibinfo {author} {\bibfnamefont
  {T.}~\bibnamefont {Liu}}, \bibinfo {author} {\bibfnamefont {J.}~\bibnamefont
  {Zhu}}, \ and\ \bibinfo {author} {\bibfnamefont {B.}~\bibnamefont {Zhang}},\
  }\href {\doibase 10.1103/PhysRevB.101.180303} {\bibfield  {journal} {\bibinfo
   {journal} {Phys. Rev. B}\ }\textbf {\bibinfo {volume} {101}},\ \bibinfo
  {pages} {180303} (\bibinfo {year} {2020})}\BibitemShut {NoStop}%
\bibitem [{\citenamefont {Bergholtz}\ \emph {et~al.}(2021)\citenamefont
  {Bergholtz}, \citenamefont {Budich},\ and\ \citenamefont
  {Kunst}}]{Bergholtz21}%
  \BibitemOpen
  \bibfield  {author} {\bibinfo {author} {\bibfnamefont {E.~J.}\ \bibnamefont
  {Bergholtz}}, \bibinfo {author} {\bibfnamefont {J.~C.}\ \bibnamefont
  {Budich}}, \ and\ \bibinfo {author} {\bibfnamefont {F.~K.}\ \bibnamefont
  {Kunst}},\ }\href {\doibase 10.1103/RevModPhys.93.015005} {\bibfield
  {journal} {\bibinfo  {journal} {Rev. Mod. Phys.}\ }\textbf {\bibinfo {volume}
  {93}},\ \bibinfo {pages} {015005} (\bibinfo {year} {2021})}\BibitemShut
  {NoStop}%
\bibitem [{\citenamefont {Ghatak}\ and\ \citenamefont
  {Das}(2019)}]{ghatak2019new}%
  \BibitemOpen
  \bibfield  {author} {\bibinfo {author} {\bibfnamefont {A.}~\bibnamefont
  {Ghatak}}\ and\ \bibinfo {author} {\bibfnamefont {T.}~\bibnamefont {Das}},\
  }\href@noop {} {\bibfield  {journal} {\bibinfo  {journal} {Journal of
  Physics: Condensed Matter}\ }\textbf {\bibinfo {volume} {31}},\ \bibinfo
  {pages} {263001} (\bibinfo {year} {2019})}\BibitemShut {NoStop}%
\bibitem [{\citenamefont {Ashida}\ \emph {et~al.}(2020)\citenamefont {Ashida},
  \citenamefont {Gong},\ and\ \citenamefont {Ueda}}]{ashida2020non}%
  \BibitemOpen
  \bibfield  {author} {\bibinfo {author} {\bibfnamefont {Y.}~\bibnamefont
  {Ashida}}, \bibinfo {author} {\bibfnamefont {Z.}~\bibnamefont {Gong}}, \ and\
  \bibinfo {author} {\bibfnamefont {M.}~\bibnamefont {Ueda}},\ }\href@noop {}
  {\bibfield  {journal} {\bibinfo  {journal} {Advances in Physics}\ }\textbf
  {\bibinfo {volume} {69}},\ \bibinfo {pages} {249} (\bibinfo {year}
  {2020})}\BibitemShut {NoStop}%
\bibitem [{\citenamefont {Kawabata}\ \emph {et~al.}(2019)\citenamefont
  {Kawabata}, \citenamefont {Shiozaki}, \citenamefont {Ueda},\ and\
  \citenamefont {Sato}}]{Kawabata19}%
  \BibitemOpen
  \bibfield  {author} {\bibinfo {author} {\bibfnamefont {K.}~\bibnamefont
  {Kawabata}}, \bibinfo {author} {\bibfnamefont {K.}~\bibnamefont {Shiozaki}},
  \bibinfo {author} {\bibfnamefont {M.}~\bibnamefont {Ueda}}, \ and\ \bibinfo
  {author} {\bibfnamefont {M.}~\bibnamefont {Sato}},\ }\href {\doibase
  10.1103/PhysRevX.9.041015} {\bibfield  {journal} {\bibinfo  {journal} {Phys.
  Rev. X}\ }\textbf {\bibinfo {volume} {9}},\ \bibinfo {pages} {041015}
  (\bibinfo {year} {2019})}\BibitemShut {NoStop}%
\bibitem [{\citenamefont {Zhou}\ \emph {et~al.}(2018)\citenamefont {Zhou},
  \citenamefont {Wang}, \citenamefont {Wang},\ and\ \citenamefont
  {Gong}}]{Zhou1}%
  \BibitemOpen
  \bibfield  {author} {\bibinfo {author} {\bibfnamefont {L.}~\bibnamefont
  {Zhou}}, \bibinfo {author} {\bibfnamefont {Q.-h.}\ \bibnamefont {Wang}},
  \bibinfo {author} {\bibfnamefont {H.}~\bibnamefont {Wang}}, \ and\ \bibinfo
  {author} {\bibfnamefont {J.}~\bibnamefont {Gong}},\ }\href {\doibase
  10.1103/PhysRevA.98.022129} {\bibfield  {journal} {\bibinfo  {journal} {Phys.
  Rev. A}\ }\textbf {\bibinfo {volume} {98}},\ \bibinfo {pages} {022129}
  (\bibinfo {year} {2018})}\BibitemShut {NoStop}%
\bibitem [{\citenamefont {Zhou}\ and\ \citenamefont
  {Du}(2021{\natexlab{b}})}]{Zhou_2021}%
  \BibitemOpen
  \bibfield  {author} {\bibinfo {author} {\bibfnamefont {L.}~\bibnamefont
  {Zhou}}\ and\ \bibinfo {author} {\bibfnamefont {Q.}~\bibnamefont {Du}},\
  }\href {\doibase 10.1088/1367-2630/ac0574} {\bibfield  {journal} {\bibinfo
  {journal} {New Journal of Physics}\ }\textbf {\bibinfo {volume} {23}},\
  \bibinfo {pages} {063041} (\bibinfo {year} {2021}{\natexlab{b}})}\BibitemShut
  {NoStop}%
\bibitem [{\citenamefont {Naji}\ \emph {et~al.}(2022)\citenamefont {Naji},
  \citenamefont {Jafari}, \citenamefont {Jafari},\ and\ \citenamefont
  {Akbari}}]{PhysRevA.105.022220}%
  \BibitemOpen
  \bibfield  {author} {\bibinfo {author} {\bibfnamefont {J.}~\bibnamefont
  {Naji}}, \bibinfo {author} {\bibfnamefont {M.}~\bibnamefont {Jafari}},
  \bibinfo {author} {\bibfnamefont {R.}~\bibnamefont {Jafari}}, \ and\ \bibinfo
  {author} {\bibfnamefont {A.}~\bibnamefont {Akbari}},\ }\href {\doibase
  10.1103/PhysRevA.105.022220} {\bibfield  {journal} {\bibinfo  {journal}
  {Phys. Rev. A}\ }\textbf {\bibinfo {volume} {105}},\ \bibinfo {pages}
  {022220} (\bibinfo {year} {2022})}\BibitemShut {NoStop}%
\bibitem [{\citenamefont {Hamazaki}(2021)}]{Hamazaki2021}%
  \BibitemOpen
  \bibfield  {author} {\bibinfo {author} {\bibfnamefont {R.}~\bibnamefont
  {Hamazaki}},\ }\href {\doibase 10.1038/s41467-021-25355-3} {\bibfield
  {journal} {\bibinfo  {journal} {Nature Communications}\ }\textbf {\bibinfo
  {volume} {12}},\ \bibinfo {pages} {5108} (\bibinfo {year}
  {2021})}\BibitemShut {NoStop}%
\bibitem [{\citenamefont {Schmitt}\ and\ \citenamefont
  {Kehrein}(2015{\natexlab{b}})}]{PhysRevB.92.075114}%
  \BibitemOpen
  \bibfield  {author} {\bibinfo {author} {\bibfnamefont {M.}~\bibnamefont
  {Schmitt}}\ and\ \bibinfo {author} {\bibfnamefont {S.}~\bibnamefont
  {Kehrein}},\ }\href {\doibase 10.1103/PhysRevB.92.075114} {\bibfield
  {journal} {\bibinfo  {journal} {Phys. Rev. B}\ }\textbf {\bibinfo {volume}
  {92}},\ \bibinfo {pages} {075114} (\bibinfo {year}
  {2015}{\natexlab{b}})}\BibitemShut {NoStop}%
\bibitem [{\citenamefont {Jafari}(2019)}]{Jafari2019}%
  \BibitemOpen
  \bibfield  {author} {\bibinfo {author} {\bibfnamefont {R.}~\bibnamefont
  {Jafari}},\ }\href {\doibase 10.1038/s41598-019-39595-3} {\bibfield
  {journal} {\bibinfo  {journal} {Scientific Reports}\ }\textbf {\bibinfo
  {volume} {9}},\ \bibinfo {pages} {2871} (\bibinfo {year} {2019})}\BibitemShut
  {NoStop}%
\bibitem [{\citenamefont {DeGottardi}\ \emph
  {et~al.}(2013{\natexlab{a}})\citenamefont {DeGottardi}, \citenamefont {Sen},\
  and\ \citenamefont {Vishveshwara}}]{DeGo1}%
  \BibitemOpen
  \bibfield  {author} {\bibinfo {author} {\bibfnamefont {W.}~\bibnamefont
  {DeGottardi}}, \bibinfo {author} {\bibfnamefont {D.}~\bibnamefont {Sen}}, \
  and\ \bibinfo {author} {\bibfnamefont {S.}~\bibnamefont {Vishveshwara}},\
  }\href {\doibase 10.1103/PhysRevLett.110.146404} {\bibfield  {journal}
  {\bibinfo  {journal} {Phys. Rev. Lett.}\ }\textbf {\bibinfo {volume} {110}},\
  \bibinfo {pages} {146404} (\bibinfo {year} {2013}{\natexlab{a}})}\BibitemShut
  {NoStop}%
\bibitem [{\citenamefont {DeGottardi}\ \emph
  {et~al.}(2013{\natexlab{b}})\citenamefont {DeGottardi}, \citenamefont
  {Thakurathi}, \citenamefont {Vishveshwara},\ and\ \citenamefont
  {Sen}}]{Manisha1}%
  \BibitemOpen
  \bibfield  {author} {\bibinfo {author} {\bibfnamefont {W.}~\bibnamefont
  {DeGottardi}}, \bibinfo {author} {\bibfnamefont {M.}~\bibnamefont
  {Thakurathi}}, \bibinfo {author} {\bibfnamefont {S.}~\bibnamefont
  {Vishveshwara}}, \ and\ \bibinfo {author} {\bibfnamefont {D.}~\bibnamefont
  {Sen}},\ }\href {\doibase 10.1103/PhysRevB.88.165111} {\bibfield  {journal}
  {\bibinfo  {journal} {Phys. Rev. B}\ }\textbf {\bibinfo {volume} {88}},\
  \bibinfo {pages} {165111} (\bibinfo {year} {2013}{\natexlab{b}})}\BibitemShut
  {NoStop}%
\bibitem [{\citenamefont {Rajak}\ \emph {et~al.}(2014)\citenamefont {Rajak},
  \citenamefont {Nag},\ and\ \citenamefont {Dutta}}]{Rajak1}%
  \BibitemOpen
  \bibfield  {author} {\bibinfo {author} {\bibfnamefont {A.}~\bibnamefont
  {Rajak}}, \bibinfo {author} {\bibfnamefont {T.}~\bibnamefont {Nag}}, \ and\
  \bibinfo {author} {\bibfnamefont {A.}~\bibnamefont {Dutta}},\ }\href
  {\doibase 10.1103/PhysRevE.90.042107} {\bibfield  {journal} {\bibinfo
  {journal} {Phys. Rev. E}\ }\textbf {\bibinfo {volume} {90}},\ \bibinfo
  {pages} {042107} (\bibinfo {year} {2014})}\BibitemShut {NoStop}%
\bibitem [{\citenamefont {Kitaev}(2001)}]{kitaev2001unpaired}%
  \BibitemOpen
  \bibfield  {author} {\bibinfo {author} {\bibfnamefont {A.~Y.}\ \bibnamefont
  {Kitaev}},\ }\href@noop {} {\bibfield  {journal} {\bibinfo  {journal}
  {Physics-uspekhi}\ }\textbf {\bibinfo {volume} {44}},\ \bibinfo {pages} {131}
  (\bibinfo {year} {2001})}\BibitemShut {NoStop}%
\bibitem [{\citenamefont {Gong}\ and\ \citenamefont {Wang}(2018)}]{Gong1}%
  \BibitemOpen
  \bibfield  {author} {\bibinfo {author} {\bibfnamefont {J.}~\bibnamefont
  {Gong}}\ and\ \bibinfo {author} {\bibfnamefont {Q.-h.}\ \bibnamefont
  {Wang}},\ }\href {\doibase 10.1103/PhysRevA.97.052126} {\bibfield  {journal}
  {\bibinfo  {journal} {Phys. Rev. A}\ }\textbf {\bibinfo {volume} {97}},\
  \bibinfo {pages} {052126} (\bibinfo {year} {2018})}\BibitemShut {NoStop}%
\end{thebibliography}%

\end{document}